\documentclass[11pt]{article}
\usepackage{lineno}
\usepackage{multirow}
\usepackage{amsmath}
\usepackage{float}
\usepackage{graphicx} % Required for inserting images
\usepackage{amssymb,amscd,amsmath,amsthm,color}
\usepackage[T1]{fontenc} % if needed
\usepackage{subcaption}
\usepackage{enumitem}
\usepackage{mathtools}
\usepackage{tcolorbox}
\usepackage{verbatim}
\usepackage{mathrsfs}
\usepackage{physics}
\usepackage{mathtools}
\usepackage{makecell}
\usepackage{slashed}
\usepackage{longtable}
\usepackage{jheppub}
\usepackage{enumerate}
\usepackage{hyperref}
\usepackage{ulem}

\numberwithin{equation}{section}

\def\NN{\mathcal{N}}

\def\<{\langle}
\def\>{\rangle}

\def\be{\begin{equation}}
\def\ee{\end{equation}}

\def\to{\rightarrow}

\makeatletter
\g@addto@macro\bfseries{\boldmath}\g@addto@macro\bfseries{\boldmath}
\makeatother
\title{\boldmath Supersymmetry and Attractors in $\mathcal{N}=4$ Supergravity: The Superconformal Approach}% Force line breaks with \\

\author{Abhinava Bhattacharjee,}
\emailAdd{abhinava19@iisertvm.ac.in}
\affiliation{School of Physics, Indian Institute of Science Education and Research Thiruvananthapuram,\\
Thiruvananthapuram 695551, India}
\author{Bindusar Sahoo}
\emailAdd{bsahoo@iisertvm.ac.in}

\abstract{
In this paper, we study the attractor mechanism for extremal, spherically symmetric black holes in pure, untruncated, $\mathcal{N}=4$ Poincaré supergravity, which we demonstrate numerically. We further study the supersymmetries preserved by these attractor solutions by focussing specifically on the constant moduli solutions and show that they always preserve 1/4$^{th}$ of the total supersymmetries. We also give an argument that even the attractor solutions with a ``non-constant'' moduli would preserve 1/4$^{th}$ of the total supersymmetries. This would mean that in pure $\NN=4$ supergravity there exist no attractor solutions which are non-supersymmetric although they could in-principle exist in a matter coupled theory. We use the framework of conformal supergravity in our analysis, which is a manifestly off-shell framework and considerably simplifies the Killing spinor analysis.}

\makeatletter
\gdef\@fpheader{}
\makeatother

\begin{document}
\allowdisplaybreaks
\maketitle
\flushbottom
\section{Introduction}\label{sec:level1}

For theories of gravity coupled to scalar fields, it has been seen that for extremal black hole solutions the scalar fields of the theory take fixed values at the black hole horizon, irrespective of their value at asymptotic infinity. This phenomenon was originally discovered for supersymmetric extremal black holes in the theories of $\mathcal{N}$-extended supergravity in four dimensions \cite{Ferrara:1995ih, Ferrara:1996dd, Ferrara:1996um}, and is known as ``attractor behaviour'' which has been the cornerstone in understanding black hole thermodynamics \cite{Bardeen:1973gs, Bekenstein:1973ur, Hawking:1975vcx} in supergravity and string theories.

$\mathcal{N}$-extended supergravity theories are extensions of the general theory of relativity that are invariant under $\mathcal{N}$-local supersymmetry transformations, in addition to diffeomorphism and local Lorentz symmetry. A black hole solution in a supergravity theory, which is a purely bosonic solution, is said to preserve some supersymmetry if it possesses globally defined Killing spinors. The existence of these Killing spinors is ensured by the vanishing supersymmetry transformation of the fermions. Schematically 
\begin{align*}
    \delta_{\epsilon} F=\partial \epsilon+ \epsilon f(B)=0\;,
\end{align*}
where the $\partial \epsilon$ term is present only when $F$ is the gravitino field. This is known as the Killing spinor equation, where $F$ and $B$ schematically denote the fermion and boson content of the supergravity theory respectively. Given a bosonic solution $B$, one can find the Killing spinor $\epsilon$ when certain integrability conditions are satisfied. Conversely, assuming the existence of a Killing spinor imposes non-trivial constraints on the bosonic fields, thereby restricting the allowed backgrounds. This approach was employed in \cite{LopesCardoso:2000qm, LopesCardoso:1998tkj} to find the maximally supersymmetric solutions in $\mathcal{N}=2$ supergravity using conformal supergravity techniques and it was found that the static extremal black holes interpolate between two maximally supersymmetric solutions: $AdS_2\times S^2$ near horizon geometry and flat spacetime at asymptotic infinity. The near-horizon configuration is completely fixed in terms of the charges due to the so-called `attractor equations'. Also known as stabilization equations, these relations determine the values of the scalar (moduli) fields of $\mathcal{N}=2$ supergravity solely in terms of the charges at the horizon. These equations played a crucial role in the derivation of the entropy formula of extremal black holes in higher-derivative $\mathcal{N}=2$ supergravity \cite{LopesCardoso:1998tkj, LopesCardoso:2000qm}, which was applied to successfully match the macroscopic black hole entropy with the microscopic counting in a wide class of $\mathcal{N}=2$ string compactifications. Over the years, the attractor mechanism has been reformulated in several ways. Notable developments include the BPS entropy function for supersymmetric black holes in $\mathcal{N}=2$ supergravity \cite{LopesCardoso:2006ugz}, as well as Sen's entropy function formalism, which applies to extremal black holes in general higher-derivative theories of gravity \cite{Sen:2005wa, Sen:2007qy}.

None of the above approaches provides a manifest description of the radial evolution of the bulk scalar fields and their dynamical flow toward the attractor values at the horizon. In contrast, the black hole potential approach developed in \cite{Ferrara:1997tw} makes the attractor mechanism explicit. Within this formalism, the radial evolution of the scalars is governed by the black hole potential $V_{BH}(\phi, Q)$, and the attractor values are determined by the extremization condition
\begin{equation}
\partial_{\phi} V_{BH}(\phi, Q) = 0 \, .
\end{equation}

The Bekenstein-Hawking entropy is then given by the extremum value of the black hole potential.
\begin{align*}
    S_{BH}=V_{BH}(\phi^{*}(Q), Q)
\end{align*}
It was subsequently realised in \cite{Ferrara:1997tw}, that attractor behaviour is not intrinsically tied to supersymmetry. Extremal black holes that do not preserve supersymmetry may nevertheless exhibit fixed-point behaviour for the scalar fields, with the attractor mechanism arising solely due to extremality, and the fixed points are essentially the extremum of the black hole potential. Such non-supersymmetric attractors have been studied predominantly in $\mathcal{N}=2$ supergravity \cite{Goldstein:2005hq, Kallosh:2006bt, kallosh2006bpsnonbpsblackholes, Sahoo:2006rp, Gimon:2007mh, LopesCardoso:2007qid, Ceresole:2007wx, Ceresole:2009vp, Cerchiai:2009pi} and type II string theory and $M$-theory compactified on Calabi-Yau threefold \cite{Tripathy:2005qp, Dabholkar:2006tb}.

The attractor mechanism in $\mathcal{N}=4$ supergravity was analyzed in \cite{Ferrara:1996um, Ferrara:1996dd} within the symplectic framework of extended ($\mathcal{N}>2$) supergravity, as formulated in \cite{Andrianopoli:1996wf, Andrianopoli:1996ve, Andrianopoli:1997xf}. The explicit supersymmetry analysis has been studied for stationary axion-dilaton solutions and dilaton black holes mostly in  the $U(1)\times U(1)$ truncated sector of $\mathcal{N}=4$ supergravity \cite{Kallosh:1992ii, Bergshoeff:1996gg, Tod:1995jf,Kallosh:1994ba, Kallosh:1993yg}.

In this work, we revisit the classic attractor mechanism for extremal, spherically symmetric black holes in pure, untruncated, $\mathcal{N}=4$ two-derivative Poincaré supergravity using the black hole potential approach. Following the approach in \cite{Goldstein:2005hq}, we first consider extremal black hole solutions where scalar fields are fixed to their attractor values in the full spacetime geometry, which we refer to as the constant moduli solutions. Thereafter, we perform a perturbative expansion around these constant moduli solutions, and determine appropriate boundary conditions that enable a consistent numerical integration of the exact equations of motion, thereby explicitly exhibiting attractor behaviour. This gives us a family of attractor solutions that are smoothly connected to the constant moduli solution. 

Since $\mathcal{N}=4$ conformal supergravity is gauge equivalent to $\mathcal{N}=4$ Poincar{\'e} supergravity, we use the framework of conformal supergravity to study the Killing spinor equations of the full $\mathcal{N}=4$ Poincaré supergravity and thereby analyse the supersymmetry of the constant moduli solution with generic charge configuration. Our primary motivation to use the framework of conformal supergravity is the following. The framework of conformal supergravity is an off-shell approach which contains a set of auxiliary fields and naturally incorporates higher derivative corrections via the field equations of the auxiliary fields. The presence of the auxiliary fields simplifies the supersymmetry transformation rules and makes the Killing spinor analysis somewhat simpler to handle. Although, we have not incorporated higher derivative corrections in our present analysis, our techniques can be generalized in a straightforward way to include higher derivative corrections.

Conformal supergravity \cite{Fradkin:1985am, Bergshoeff:1980is, deWit:1979dzm} is an extension of Poincaré supergravity where the local symmetries such as diffeomorphism, local Loerntz symmetry and the ordinary supersymmetry (often referred to as $Q$-susy) are augmented with additional local symmetries such as the dilatation, special conformal transformation and special supersymmetry (often referred to as $S$-susy) as well as R-symmetries (in the case of extended supergravity theories). The superconformal algebra can be realised off-shell by the inclusion of a set of auxiliary fields, providing a powerful and systematic way to incorporate higher-derivative corrections to the Poincaré supergravity action. The basic multiplet of a conformal supergravity theory is called the Weyl multiplet. This contains all the independent gauge fields arising from the superconformal algebra, along with other matter and auxiliary fields needed for the off-shell closure of the algebra. Apart from the Weyl multiplets, matter multiplets also play an important role as it is the matter coupled conformal supergravity that is gauge equivalent to Poincar{\'e} supergravity. In order to go from conformal to Poincar{\'e} supergravity, one needs to couple a minimum number of matter multiplets to conformal supergravity which provides the necessary degrees of freedom to compensate for the extra symmetries present in conformal supergravity. Other than these compensating matter multiplets, one can also couple extra matter multiplets to conformal supergravity in order to arrive at matter coupled Poincar{\'e} supergravity. In order to obtain the Poincaré supergravity action from the matter coupled conformal supergravity action, one applies gauge fixing conditions on the compensating multiplets and eliminates the auxiliary fields using their equations of motion. Whether one obtains a Poincar{\'e} supergravity action with higher derivatives depends on whether one has included the pure conformal supergravity action involving just the Weyl multiplet before employing the above mentioned procedures. The presence of the pure conformal supergravity action would change the auxiliary field equations which can be eliminated in an order by order fashion thereby giving Poincar{\'e} supergravity action as an expansion in derivatives \cite{Ciceri2025}. The gauge equivalence principle can also be employed at the level of the transformation rules in order to obtain super-Poincar{\'e} transformation from the superconformal transformation. The gauge fixing conditions necessarily requires to modify the superconformal transformation rules by some field dependent gauge transformations. Further, one may also eliminate the auxiliary fields by their equations of motion to obtain the ``on-shell'' super-Poincar{\'e} transformations that will only close upon using the equations of motion.

The plan of the paper is as follows. In section \ref{N=4consugra}, we briefly describe the elements of $\mathcal{N}=4$ conformal supergravity that is relevant for our work. In section \ref{N=4ponsugra}, we discuss how we can obtain the pure $\mathcal{N}=4$ Poincaré supergravity action from matter coupled $\NN=4$ conformal supergravity. In section \ref{1daction}, we consider spherically symmetric ansatz of the underlying fields of $\NN=4$ Poincar{\'e} supergravity and reduce the action to a one-dimensional effective action. The equations of motion arising from this one-dimensional effective action will be instrumental in finding the spherically symmetric solutions to $\NN=4$ Poincar{\'e} supergravity. In section \ref{frozen}, we numerically solve these equations of motion using appropriate boundary conditions and demonstrate the attractor behaviour. In section \ref{susyconstant}, we analyze the Killing spinor equations and explicitly show that the constant moduli solutions are $1/4$-BPS for a generic charge configuration. We have also included couple of appendices. In appendix-\ref{conv}, we give our notations and conventions. In appendix-\ref{convergence}, we show the convergence of the perturbative series solutions obtained in section-\ref{pertsol}.
\section{N=4 Conformal Supergravity} \label{N=4consugra}
In this section, we briefly describe the elements of $\mathcal{N}=4$ conformal supergravity developed in \cite{deRoo:1984zyh, Ciceri:2015qpa,Butter:2016mtk,Butter:2019edc}. The key multiplets in $\mathcal{N}=4$ conformal supergravity are the Weyl multiplet and the vector multiplet.

The Weyl multiplet contains the following fields: The independent gauge fields are the vierbein $e_{\mu}{}^{a}$, the dilatation gauge field $b_{\mu}$, the $SU(4)$ R-symmetry gauge field $V_{\mu}{}^{i}{}_{j}$, the $Q$-supersymmetry (or ordinary supersymmetry) gauge field $\psi_{\mu}{}^{i}$ (gravitini) and the dependent gauge fields are the spin connection $\omega_{\mu}{}^{ab}$, the chiral $U(1)$ gauge field $a_{\mu}$, $S$-supersymmetry (or special supersymmetry) gauge field $\phi_{\mu}{}^{i}$ and the gauge field associated with special conformal transformation $f_{\mu}{}^{a}$. 

It also contains several matter fields (both dynamical as well as auxiliary): The bosonic matter fields are the following. There is a real anti-symmetric Lorentz tensor whose anti-self dual part is denoted as $T_{ab}{}^{ij}$ transforming in the $\textbf{6}$ representation of $SU(4)$ and self-dual part is denoted as $T_{abij}$ transforming in the ${\mathbf{\overline{6}}}$ representation of $SU(4)$. There are complex scalars $E_{ij}$ transforming in the $\mathbf{\overline{10}}$ representation of $SU(4)$. There are pseudo-real scalars $D^{ij}{}_{kl}$ transforming in $\textbf{20}'$ representation of $SU(4)$ satisfying the following pseudo-reality condition:
\begin{align}
\left(D^{ij}{}_{kl}\right)^*\equiv D^{kl}{}_{ij}=\frac14 \varepsilon^{klmn}\varepsilon_{ijpq}D^{pq}{}_{mn}
\end{align}
There is a doublet of complex scalars $\phi_{\alpha}$ ($\alpha=1,2$) charged under a local $U(1)$ and transforming in the fundamental of a global $SU(1,1)$ and is subject to a $SU(1,1)\times U(1)$ invariant constraint 
\begin{align}
    \phi^{\alpha}\phi_{\alpha}=1,~\phi^{\alpha}=\eta^{\alpha\beta}(\phi_{\beta})^{*}\;,
\end{align}
where $\eta^{\alpha\beta}=\text{diag}(1,-1)$ is the $SU(1,1)$ metric. The scalars $\phi_{\alpha}$ parametrise an $SU(1,1)/U(1)$ coset manifold and hence are often referred to as the `coset scalars'. The fermionic matter fields comprise of the following. There is a quartet of Majorana spinor whose left chiral part is denoted as $\Lambda_{i}$ transforming in the $\mathbf{\overline{4}}$ representation of the $SU(4)$ R-symmetry. Further there is a set of Majorana spinor whose left chiral part is denoted as $\chi^{ij}{}_{k}$ transforming in the ${\textbf{20}}$ representation of $SU(4)$ R-symmetry. Here $i,j=1,\cdots,4$ denote the $SU4)$ R-symmetry indices. We also follow a chiral notation where the raising and lowering of the R-symmetry indices is done via complex conjugation. For example:
\begin{align}
    E^{ij}&=(E_{ij})^*\nonumber \;, \\
    \Lambda^{i}&=i\gamma^0C^{-1}(\Lambda_i)^*\;,
\end{align}
where in the second line $C$ is the charge-conjugation matrix. The complex conjugation operation in the second line flips the chirality of the fermion and the relation follows from the Majorana condition.

The $Q$ and $S$ supersymmetry transformations are parametrized by the parameters $\epsilon^{i}$ and $\eta^{i}$. In this paper, the relevant supersymmetry transformations are the one that acts on the fermions and are given as 

\begin{align}\label{trans}
   \delta \psi_\mu{}^i=\,&2\,\mathscr{D}_\mu\epsilon^i
   -\tfrac12\gamma^{ab}T_{ab}{\!}^{ij}\gamma_\mu\epsilon_j
  -\gamma_\mu\eta^i \,,\nonumber\\ 
 \delta \chi^{ij}{\!}_k=\,&-\tfrac12\gamma^{ab}
   \slashed{D}T_{ab}{\!}^{ij}\epsilon_k
   -\gamma^{ab} R(V)_{ab}{}^{[i}{}_k \,\epsilon^{j]}
  \nonumber\\
  &-\tfrac12\varepsilon^{ijlm}\,\slashed{D} E_{kl}\,\epsilon_m +D^{ij}{\!}_{kl}\, \epsilon^l
  \nonumber\\
  & -\tfrac1{6}\varepsilon_{klmn}E^{l[i}\gamma^{ab}\big[T_{ab}{\!}^{j]n}\epsilon^m+T_{ab}{\!}^{mn}\epsilon^{j]}\big]
  \nonumber\\
  & +\tfrac12 E_{kl}\,E^{l[i}\,\epsilon^{j]}
   -\tfrac12 \varepsilon^{ijlm}\bar{\slashed{P}}\gamma_{ab}
   T^{ab}{\!}_{kl}\,\epsilon_m\nonumber\\
     \delta \Lambda_i=\,&-2\,\bar{\slashed{P}}\epsilon_i
   +E_{ij}\epsilon^j+\tfrac12\varepsilon_{ijkl}\,T_{bc}{\!}^{kl} 
   \gamma^{bc} \,\epsilon^j \,,
\end{align}
where we have omitted the fermionic terms in the transformation rules and $\mathscr{D}_\mu\epsilon^i$ is defined as
 \begin{align}\label{cdepsilon}
         \mathscr{D}_{\mu}\epsilon^{i}=\,&\Big[\partial_{\mu}-\frac{1}{4}\omega_{\mu}{}^{ab}\gamma_{ab}+\frac{1}{2}(b_{\mu}+\operatorname{i}a_{\mu})\Big]\epsilon^{i}-V_{\mu}{}^{i}{}_{j}\epsilon^{j}
 \end{align}
 We will use the same symbol $\mathscr{D}_{\mu}$, when acting on any covariant object, to denote a derivative that is covariant with respect to all bosonic symmetries except special conformal transformation, whereas in contrast, the derivative $D_\mu$ appearing in the above transformation laws is the fully supercovariant derivative. $P_{\mu}$ is the Maurer-Cartan one-from associated with $SU(1,1)$ and $a_{\mu}$ is the composite $U(1)$ gauge field defined as:
\begin{align}
    P_{\mu} &= \varepsilon_{\alpha\beta} \phi^{\alpha} D_{\mu} \phi^{\beta} \notag\\
    \Bar{P}_{\mu} &= -\varepsilon^{\alpha\beta} \phi_{\alpha} D_{\mu} \phi_{\beta} \notag\\
    a_{\mu} &= i \phi^\alpha \partial_{\mu} \phi_{\alpha} +\frac{i}{4} \Bar{\Lambda}^i \gamma_{\mu} \Lambda_{i}
\end{align}
The complete transformation rules for the Weyl multiplet can be found in \cite{deRoo:1984zyh, Ciceri:2015qpa,Butter:2016mtk,Butter:2019edc}.

A single abelian $\mathcal{N}=4$ vector multiplet contains a gauge field $A_{\mu}$, scalar fields $\phi_{ij}$, and spin-1/2 fermion $\psi_{i}$ (gaugino). Their properties are summarized in Table \ref{tab:vector}.  

\begin{table}
\centering
    \begin{tabular}{|c|c|c|c|c|}
    \hline
         Field& Properties& $SU(4)$ &$w$ &$c$ \\
         \hline
         $A_{\mu}$ & Gauge field & $\textbf{1}$ &0&0\\ \hline
         $\psi_i$& $\gamma_5 \psi_i=-\psi_i$ & $\overline{\textbf{4}}$ & $3/2$& $-1/2$\\  \hline
         $\phi_{ij}$& \makecell{Pseudo-real: \\ \(\phi^{ij}\equiv(\phi_{ij})^*=-\tfrac{1}{2}\varepsilon^{ijkl}\phi_{kl}\)} & $\textbf{6}$ & 1 &0\\
         \hline
    \end{tabular}
    \caption{$\mathcal{N}=4$ Vector multiplet}
    \label{tab:vector}
\end{table}

In order to obtain $\NN=4$ Poincar{\'e} supergravity coupled to $n_v$ vector multiplets one needs to couple $6+n_v$ vector multiplets to $\NN=4$ conformal supergravity, where the six extra vector multiplets are necessary to compensate for the extra symmetries present in conformal supergravity. In our work, we will need the supersymmetry transformations (both $Q$ as well as $S$) for the $(6+n_v)$ gauginos, labeled by $I=1,2,\cdots,(6+n_v)$, present in these vector multiplets. They are given as
\begin{align}\label{transvec}
    \delta \psi_{i}^{I}=\,&-\frac{1}{2\Phi_{+}}\gamma^{ab}\epsilon_{i}\Big({F}_{ab}^{I +}+\Phi^*_{+}T_{ab ij}\phi^{I ij}\Big)-2\slashed{D}\phi_{ij}^{I}\epsilon^j+E_{ij}\phi^{I jk}\epsilon_{k} -2\phi_{ij}^{I}\eta^{j}\,,
    \end{align}
    where,
    \begin{equation}
        \Phi_{\pm}=\phi^1\pm\phi^2, ~~~\Phi_{\pm}^*=\phi_1\mp\phi_2\;,
    \end{equation}
and $\hat{F}_{\mu\nu}^{I+}$ is the self-dual part of the supercovariant field strength\footnote{See appendix-\ref{conv} for the definition of a self-dual tensor in our convention.} associated with the gauge fields $A_{\mu}^I$ given as:
    \begin{equation}
{F}_{\mu\nu}^{I}=2\partial_{[\mu}A_{\nu]}^{I}
    \end{equation}
For our work, the bosonic part of the action for $(6+n_v)$ vector multiplets in the background of the Weyl multiplets is relevant, which was derived in \cite{deRoo:1984zyh} and is given as: 
\begin{align}\label{VM-CSG}
    e^{-1}\mathcal{L}_{V}=\,&-\frac{\Phi_{-}}{4\Phi_{+}}F_{ab}^{+I}F^{abJ}\eta_{IJ}+\frac{1}{4}\phi_{ij}^{I}D^2\phi^{Jij}\eta_{IJ}+\frac{1}{8}\phi^{I}_{ij}\phi^{Jkl}D^{ij}{}_{kl}\eta_{IJ}\nonumber\\
    \,& -\frac{1}{48}\phi^{I}_{ij}\phi^{Jij}\eta_{IJ}[E^{kl}E_{kl}-4P_{a}\bar{P}^{a}]-\frac{\Phi_{+}^{*}}{2\Phi_{+}}T^{ab}{}_{ij}T_{abkl}\phi^{Iij}\phi^{Jkl}\eta_{IJ}\nonumber\\
    \,&-\frac{1}{\Phi_{+}}F_{ab}^{I}T^{ab}{}_{ij}\phi^{Jij}\eta_{IJ}+\text{h.c.}
\end{align}
where 
\begin{align}
          \eta= \,&diag(\underbrace{-1,\dots ,-1}_\text{6}, \underbrace{+1,\dots ,+1}_\text{$n_v$})
\end{align}
is the $SO(6,n_{v})$ invariant metric and $D^2\phi^{Iij}\equiv D_{a}D^{a}\phi^{Iij}$.
\section{Pure N=4 Poincar\'e Supergravity}\label{N=4ponsugra}
In this section, we briefly describe the construction of pure Poincaré supergravity action.
As mentioned in the introduction, the Poincaré supergravity action is obtained by gauge fixing the extra symmetries and substituting the auxiliary fields with their equations of motion \cite{deRoo:1984zyh, Ciceri2025}. First, we present the gauge fixing conditions below in Table \ref{tab:gaugecon}. We have refrained from giving an explicit $SU(4)$ gauge fixing condition. It turns out that as far as the bosonic action is concerned, it can be expressed completely in terms of the $SU(4)$ invariant object $\phi^{Iij}\phi^{J}_{ij}$ which is a constant in the case of pure $\NN=4$ supergavity given by \eqref{purephi}. However, later when we analyse the supersymmetry of solutions in section-\ref{susyconstant}, we will need an explicit $SU(4)$ gauge fixing condition, which we will discuss in details in that section. %For $n_v=0$, we have $6$ vector multiplets which contains $6$ vector fields $A_{\mu}^{I}$, $36$ scalars $\phi^{I}_{ij}$ and $6$ gauginos $\psi_{i}^{I}$. These fields can be used to gauge fix the extra symmetries as follows:

 \begin{table}[H]
\small
    \centering
     \begin{tabular}{|c|c|c|}
     \hline
          Symmetries& Gauge fixing conditions &Name \\
          \hline
       SCT& $b_{\mu}=0$& $K$-gauge\\
       \hline
       Dilatation & $\phi^{Iij}\phi_{ij}^J\eta_{IJ}=-\frac{6}{\kappa^2}$ &$D$-gauge\\ \hline
         $S$-supersymmetry & $\phi^{Iij}\psi^{J}_{j}\eta_{IJ}=0$  &$S$-gauge\\
          \hline
     \end{tabular}
     \caption{Gauge fixing conditions}
     \label{tab:gaugecon}
 \end{table}
Here $\kappa=\sqrt{8\pi G}$ , where $G$ is Newton's constant. From now on, we adapt to a natural unit where $\kappa=1$.

For our purpose, we only need the bosonic auxiliary field equations, where we also omit any fermionic terms. The field $D^{ij}{}_{kl}$ appears as a Lagrange multiplier in the Lagrangian and imposes the following constraint:
\begin{align}\label{Deom}
    \left(\phi^{Iij}\phi^{J}_{kl}\eta_{IJ}\right)_{\bf{20'}}=0
\end{align}
This constraint can be combined with the $D$-gauge condition to obtain the following: 
\begin{equation}\label{dijkl}
    \phi^{I ij}\phi^{J}_{ 
 kl}\eta_{IJ}= -\delta^{i}_{[k}\delta^{j}_{l]}
\end{equation}
The $T_{ab}^{ij}$ and $E_{ij}$ field equations read respectively
\begin{align}
    T_{abij}=\,&\frac{1}{\Phi_{+}^{*}}F_{ab}^{+I}\phi_{ij}^{J}\eta_{IJ}\label{auxitab}\\
    E_{ij}=\,&0\label{eij}
\end{align}
The $SU(4)$ gauge field $V_{\mu}{}^{i}{}_{j}$ appears inside the covariant derivatives in the Lagrangian. Its equations of motion reads
\begin{align}
    V_{\mu}{}^{i}{}_{j}=\frac{1}{2}\Big(\phi^{Iij}\partial_{\mu}\phi_{jk}^{J}-\phi^{I}_{jk}\partial_{\mu}\phi^{Jik}\Big)\eta_{IJ}-\text{trace}\label{vmuij}
\end{align}
Since the field $D^{ij}{}_{kl}$ appears as a Lagrange multiplier, an explicit solution for it is not required to write down the Poincar{\'e} action. However, one can obtain the explicit expression for $D^{ij}{}_{kl}$ by doing the following. Let us take the field equations corresponding to the scalar $\phi^{Iij}$ which is given as:
\begin{align}
    D^2\phi^{Iij}\,&+\frac{1}{2}\phi^{I}_{kl}D^{kl}{}_{ij}-\frac{1}{12}\phi^{I}_{ij}(E^{kl}E_{kl}-4P_{a}\bar{P}^{a})-\bar{F}_{ab}^{I}(T^{ab}{}_{ij}-\frac{1}{2}\varepsilon_{ijkl}T^{abkl})=0
\end{align}
where 
\begin{align}
    \bar{F}_{ab}^{I}=\frac{1}{2\Phi_{+}}F_{ab}^{I}+\frac{\Phi^{*}_{+}}{\Phi_{+}}T_{abij}\phi^{Iij}+h.c.
\end{align}
We then contract the above equation with $\phi^{J}_{kl}\eta_{IJ}$ and take the projection on the $\bf{20'}$ representation of $SU(4)$ to obtain the following expression for $D^{ij}{}_{kl}$ which we present for the sake of later use.
\begin{align}
    D^{ij}{}_{kl}=2[\phi^{Jij}D^2\phi^{J}_{kl}\eta_{IJ}]_{\textbf{20}'}\label{eomdijkl} \;,
\end{align}
where we have used \eqref{dijkl} and \eqref{auxitab} to arrive at the result.
We will also need the Maxwell's equation arising from the action \eqref{VM-CSG}. It takes the following form:
\begin{align}
    D^{a}\Big(G^{I+}_{ab}+G^{I-}_{ab}\Big)=\,&0
\end{align}
where
\begin{align}
    G^{+Iab}=\frac{\Phi_{-}}{\Phi_{+}}F^{+Iab}+ \frac{2}{\Phi_{+}}T^{ab}{}_{ij}\phi^{Iij}
\end{align}
The Bianchi identity is given as:
\begin{align}
     D^{a}\Big(F^{I+}_{ab}-F^{I-}_{ab}\Big)=\,&0
\end{align}
The $SU(1,1)$ transformation acting on the coset scalars $\phi^{\alpha}$ is not a symmetry of the action \eqref{VM-CSG} but nevertheless it acts as an electromagnetic duality symmetry of the Maxwell's equation together with the Bianchi identity by acting as an $SL(2,R)$ transformation on the field strength $F^{I}_{ab}$ and the dual field strength $G^{I}_{ab}$ (For details refer \cite{deRoo:1984zyh}). For later purpose, we can combine the Maxwells equation with the Bianchi identity and the auxiliary field equations to obtain the following:
\begin{align}\label{Tabmaxwell}
    D^{a}T_{abij}=-\frac{1}{2}P^{a}T_{ab}{}^{kl}\varepsilon_{ijkl}
\end{align}
Further, $\mathcal{N}=4$ conformal supergravity is written in a language where a spurious local $U(1)$ symmetry acting on the coset scalars is present. We may choose to fix this symmetry by choosing the following triangular gauge condition.
\begin{align}
    \operatorname{Im}(\phi_1-\phi_2)=0
\end{align}
In this gauge, the coset scalars can be expressed in terms of a single complex scalar field $\tau$:
\begin{align}
    \phi_1=\frac{1}{2\sqrt{\operatorname{Im} \tau}}(1-i\tau), ~\phi_2=-\frac{1}{2\sqrt{\operatorname{Im} \tau}}(1+i\tau)
\end{align}
The Maurer-Cartan one-form $P_{\mu}$ and the composite gauge field $a_{\mu}$ in terms of $\tau$ read
\begin{subequations}
    \begin{align}
    P_{\mu}=\,&\frac{\partial_{\mu}\tau}{2i\operatorname{Im}\tau}\label{pmutau}\\
    a_{\mu}=\,&\frac{\partial_{\mu}(\tau+\bar{\tau})}{4 \operatorname{Im}\tau}\label{taumu}
\end{align}
\end{subequations}
The complex scalar field $\tau=\phi+i \chi$ known as the axion-dilaton moduli are the physical scalar fields of the $\mathcal{N}=4$ Poincaré supergravity.

Now we are in a position to present the action for pure $\mathcal{N}=4$ Poincar\'e supergravity which corresponds to $n_v=0$. We then have $6$ vector multiplets which contains $6$ vector fields $A_{\mu}^{I}$, $36$ scalars $\phi^{I}_{ij}$ and $6$ gauginos $\psi_{i}^{I}$. In this case, all the 36 scalar can be set to constants. This is done as follows:
$21$ out of $36$ scalar fields can be fixed to constants by the equation \eqref{Deom} and the $D$-gauge condition. The remaining $15$ scalars can be fixed to a constant using local $SU(4)$ transformation. In this $SU(4)$ gauge, all the scalars are constant and satisfy 
\begin{align}\label{purephi}
    \phi^{Iij}\phi^{J}_{ij}=-\eta^{IJ}, ~~\eta^{IJ}\eta_{JK}=\delta^{I}{}_{K}
\end{align}
where 
\begin{align}
    \eta= \,&diag(\underbrace{-1,\dots ,-1}_\text{6})
\end{align}
Substituting all the auxiliary field equations and gauge fixing conditions discussed before, we obtain the action for the bosonic sector pure $\mathcal{N}=4$ Poincar\'e supergravity
\begin{align}\label{purelead}
\mathcal{L}=-\frac{1}{2}R-G_{\tau\bar{\tau}}\partial_{\mu}\bar{\tau} \partial^{\mu}\tau -f_{IJ}F_{ab}^I F^{ab J}-\tilde{f}_{IJ}F_{ab}^I \tilde{F}^{ab J}\;,
\end{align}
where
\begin{align}\label{modmetric}
    G_{\tau\bar{\tau}}=\frac{1}{{4(\text{Im}\tau)^2}}
\end{align}
is the metric in the moduli space parametrized by the coset scalars. We have also defined,
\begin{align}\label{coupling}
       f_{IJ}=-\frac{\chi}{4}\eta_{IJ},\, \tilde{f}_{IJ}=\frac{i\phi}{4}\eta_{IJ}\;.
\end{align}
Note that in this section we did not consider the pure conformal supergravity action built purely out of the Weyl multiplet fields \cite{Butter:2016mtk,Butter:2019edc}, which is a purely higher derivative action. If we do that, then the equations of motion of the auxiliary fields will no longer be algebraic and has to be eliminated order by order in derivatives. This will lead to an $\NN=4$ Poincar{\'e} supergravity action higher order in derivatives where the elimination of the auxiliary fields will result in a derivative expansion. For details, see the upcoming paper \cite{Ciceri2025}. However, for simplicity we do not consider higher derivative $\NN=4$ supergravity in this paper and restrict ourselves to \eqref{purelead}.
\section{1D Effective Action}\label{1daction}
Since we are interested in spherically symmetric solutions, it is useful to restrict ourselves to spherically symmetric ansatz for the underlying fields and obtain a 1-d effective action. This is similar to the analysis of \cite{Goldstein:2005hq}. We start by taking the following spherically symmetric ansatz for the metric:
\begin{equation}\label{metrican}
   ds^2=-a(r)^2dt^2+ a(r)^{-2}dr^2+b(r)^2d\Omega^2\;,
    \end{equation}
whose isometry group is $SO(3)$. Further, the gauge field strengths consistent with this isometry and the Bianchi identity takes the following standard form
\begin{equation}\label{fielda}
    {F^{I}= e^{I}(r)dt\wedge dr+ p^{I}\sin\theta d\theta\wedge d\phi}
\end{equation}
where $p^I$'s are constants and  $e^{I}(r)=-\partial_{r}A_t^{I}(r)$ are the electric fields. The scalar fields consistent with the $SO(3)$ isometry are restricted to depend only on the radial variable as:
\begin{equation}\label{taua}
    \tau= \tau(r)
\end{equation} 
Now, we perform a direct substitution of the ansatz \eqref{metrican},\eqref{fielda} and \eqref{taua}
 into the action \eqref{purelead} and integrate out the angular degrees of freedom. We obtain a one dimensional effective action:
 \begin{align}\label{redu}
       \mathcal{S}_{1D}=\int dr \,&\Big[1+2aba'b'+a^2b'^2 -G_{\tau\Bar{\tau}}a^2b^2\Bar{\tau}'\tau'+\nonumber\\\,&\frac{2}{b^2}(e^{I}e^{J}f_{IJ}b^4-p^{I}p^{J}f_{IJ})-4ip^{I}e^{J}\Tilde{f}_{IJ}\Big]
\end{align}
where the primes are derivative w.r.t $r$. This is a classical mechanical action with the generalized coordinates $(a, b, \tau, \Bar{\tau}, A_{t}^{I})$. The equations of motion are
\begin{subequations}
    \begin{align}
     \frac{b''}{b}&=- G_{\tau\Bar{\tau}}\Bar{\tau}'\tau'\label{bequation}\\
      -G_{\tau\Bar{\tau}}a^2 b \Bar{\tau}'\tau'+ 2 e^{I}e^{J}f_{IJ} b+ \frac{2}{b^3}f_{IJ}p^{I}p^{J} &=-(ba'^2+baa''+a a' b'+a^2 b'')\label{intermediate}\\
     (G_{\tau\Bar{\tau}}a^2 b^2 \Bar{\tau}')'-\partial  G_{\tau\Bar{\tau}}a^2b^2\Bar{\tau}'\tau'&=- 2\partial f_{IJ}e^{I}e^{J}b^2+ 2\partial f_{IJ}\frac{p^{I}p^{J}}{b^2}+4i \partial \Tilde{f}_{IJ}e^{I}p^{J} \label{taueom}\\
      (b^2 f_{IJ}e^{J}-i\Tilde{f}_{IJ}p^{J})'&= 0\label{maxredu}
\end{align}
\end{subequations}
Note that the effective Lagrangian has a $r$-translation symmetry and hence one can obtain the following conserved quantity (similar to the Hamiltonian of classical mechanics). 
\begin{align}
    \mathcal{Q}=\,& 2aba'b'+a^2b'^2-G_{\tau\Bar{\tau}}a^2b^2\Bar{\tau}'\tau'
    +\frac{2}{b^2}(e^{I}e^{J}f_{IJ}b^4+f_{IJ}p^{I}p^{J})-1\;.
\end{align}
Beside having the $r$-translation invariance, \eqref{redu} has a $r$- reparametrization invariance inherited from the diffeomorphism invariance of the original four dimensional action. Under the reparametrization $\Tilde{r}=r+f(r)$ the action varies as 
\begin{equation*}
    \delta S_{1D}=-\int \mathcal{Q} f'(r)=0\;. 
\end{equation*}
This implies 
\begin{align}\label{hc}
    \mathcal{Q}=&2aba'b'+a^2b'^2-G_{\tau\Bar{\tau}}a^2b^2\Bar{\tau}'\tau'+\frac{2}{b^2}(e^{I}e^{J}f_{IJ}b^4+f_{IJ}p^{I}p^{J})-1=0\;,
\end{align} 
which we refer to as the Hamiltonian constraint. Now combining \eqref{hc} with \eqref{intermediate}, we obtain
\begin{equation}\label{mm}
    (a^2b^2)''=2
\end{equation}
We can exactly solve Eq. \eqref{maxredu} in terms the conserved $U(1)$ charges $q_{I}$:
\begin{equation}e^{I}=\frac{f^{IJ}(q_{J}+i\Tilde{f}_{JK}p^K)}{b^2}
\end{equation}
Putting this back in the $\tau$ equations of motion \eqref{taueom} and the Hamiltonian constraint \eqref{hc}, we get-
     \begin{equation}\label{scl}
 (G_{\tau\Bar{\tau}}a^2 b^2 \Bar{\tau}')'-\partial  G_{\tau\Bar{\tau}}a^2b^2\Bar{\tau}'\tau'=\frac{2}{b^2} \frac{\partial V}{\partial \tau}
    \end{equation}
    and
    \begin{equation}\label{hcv}
        2aba'b'+a^2b'^2-G_{\tau\Bar{\tau}}a^2b^2\Bar{\tau}'\tau'+ \frac{2}{b^2}V-1=0
    \end{equation}
    where $V$ is the so-called black hole potential which is a real function of the form
    \begin{equation}\label{bp}
        V=f^{IJ}(q_{I}+i\Tilde{f}_{IK}p^K)(q_{J}+i\Tilde{f}_{JM}p^M)+ f_{IJ}p^Ip^J
    \end{equation}
    \section{Constant Moduli Solution}\label{frozen}
    In this section, we focus on one special class of solution where we set the scalar field to a constant value $\tau(r)= \tau_{0}$ for the full spacetime geometry. We can consistently do that as long as $V$ has a critical point at $\tau=\tau_{0}$  cf. \eqref{scl}. Setting the first derivative of $V$ with respect to $\tau$ to zero at $\tau=\tau_0$, we get:
\begin{align}\label{tau0sol}
 \frac{16}{(\tau_{0}-\,\Bar{\tau}_{0})^2}q_{I}\eta^{IJ}q_{J}-\,&\frac{8\Bar{\tau}_{0}}{(\tau_{0}-\Bar{\tau}_{0})^2}q_{I}p^{I}+\frac{\Bar{\tau}_{0}^2}{(\tau_{0}-\Bar{\tau}_{0})^2}p^{I}\eta_{IJ}p^{J}=0
\end{align}
Assuming $\tau_0-\bar{\tau}_0$ to be non-zero and finite, we get a quadratic equation in $\Bar{\tau}_0$ whose solution is
\begin{equation}\label{ts}
    \Bar{\tau}_{0}= \frac{4\,p.q\pm 4\,i\sqrt{q^2p^2-(p.q)^2}}{p^2}
\end{equation}
where $p^2=p^I\eta_{IJ}p^J<0, q^2=q_{I}\eta^{IJ}q_J<0$, $p.q=p^Iq_{I}$ and $p^2q^2>(p.q)^2$. Note that if $q^2p^2<(p.q)^2$, then \eqref{ts} is completely real which violates our assumption of $\tau_0-\bar{\tau}_0$ to be non-zero and hence will not solve \eqref{tau0sol}.\footnote{In this case the L.H.S of \eqref{tau0sol} is of the form $0/0$ but the numerator is a simple zero whereas the denominator would be a double zero.} Further, the moduli space metric \eqref{modmetric} diverges if $\operatorname{Im} \tau=0$. Hence we need to take $p^2q^2>(p.q)^2$.\footnote{Note that this condition will be allowed to be relaxed if we were working with matter coupled supergravity instead of pure supergravity.} Furthermore, $\tau_0$ has to lie on the upper half plane, otherwise the vector kinetic terms in the Lagrangian \eqref{purelead} would have the wrong sign. Since $p^2<0$, the allowed solution is 
\begin{equation}\label{phichisol}
    \phi_0\equiv \operatorname{Re}(\tau_0)=\frac{4\,q.p}{p^2},~~\chi_{0}\equiv \operatorname{Im}(\tau_0)=-\frac{4\,\sqrt{q^2p^2-(p.q)^2}}{p^2}
\end{equation}
Let us check the stability of the solution. In terms of $\phi$ and $\chi$ the potential \eqref{bp} takes the form,
\begin{equation}
    V(\phi,\chi)=-\frac{4\,q^2}{\chi}-\frac{\phi^2+\chi^2}{4\chi}p^2+ \frac{2\phi}{\chi}(q.p)
\end{equation}
The Hessian of $V$,
\begin{equation}
    H_{V}=\begin{pmatrix}
           \frac{\partial^2 V}{\partial \phi^2} &  \frac{\partial ^2 V}{\partial \phi \partial \chi}\\
           \\
        \frac{\partial ^2 V}{\partial \phi \partial \chi} &\frac{\partial^2 V}{\partial \chi^2}\\
    \end{pmatrix}
\end{equation}
is a positive definite matrix for the solution \eqref{phichisol} since one can explicitly check that (on the solution),
\begin{align}\label{hessian}
     &\frac{\partial^2 V}{\partial \phi^2}=\frac{\partial^2 V}{\partial \chi^2}=\frac{p^4}{8\sqrt{q^2p^2-(q.p)^2}}>0,~~~\frac{\partial ^2 V}{\partial \phi \partial \chi}=0     
\end{align}
Hence the solution \eqref{phichisol} is indeed a stable solution. Now we turn to the solution of the metric. For the constant moduli solution, equation-\eqref{bequation} gives
\begin{equation}
    b_{0}(r)=A_1r+A_2,
 \end{equation}
By exploiting the $r$-reparametrization symmetry, we can redefine the radial variable ${r}\rightarrow A_1r+A_2$, and set $b_{0}(r)=r$. In this gauge, the solution for $a(r)$  is obtained by solving equation-\eqref{mm} and is given as:
\begin{equation}
    a_{0}^2(r)=\Big(1+\frac{B_1}{r}+\frac{B_2}{r^2}\Big)
\end{equation}
Since $a_0$ appear in the $g_{tt}$ and $g_{rr}$ components of the metric \eqref{metrican}, its zeros gives us the position of the horizons as:
\begin{equation}
    r_{\pm}=\frac{-B_1\pm \sqrt{B_1^2-4B_2}}{2}
\end{equation}
Putting the solution $a_{0}$ and $b_{0}$ in the Hamiltonian constraint \eqref{hcv}, we obtain
\begin{equation}\label{rhdef}
B_2= 2V(\phi_0,\chi_0)= 4\sqrt{q^2p^2-(p.q)^2}\equiv r_{H}^2
\end{equation}
By setting $B_1=-2r_H$, we get an extremal black hole of horizon radius $r_{H}$, where the metric coefficients are given as:
\begin{equation}
   b_0(r)=r,~~a_{0}(r)=\Big(1-\frac{r_H}{r}\Big)
\end{equation}
The Bekenstein-Hawking entropy of these extremal black holes is given as:
\begin{align}
    S_{BH}=\frac A4=4\pi\sqrt{p^2q^2-(p.q)^2}
\end{align}
In this section so far, we have seen that, for the special case of constant moduli, we obtain an extremal black hole solution where the moduli and the geometry of the full solution, as well as the black hole's entropy, are completely specified in terms of the charges. In the next section, we will perturb around the constant moduli solution and see changes in the full geometry. However, the moduli and the geometry near the horizon will be completely fixed by the charges irrespective of the perturbation, and hence the entropy will remain the same, thereby exhibiting the attractor behaviour at the perturbative level. 
\section{Perturbative solutions}\label{pertsol}
We now turn on a small perturbation $\tau_{1}$ to the scalar field around the critical value $\tau_{0}$
\begin{equation}
    \tau=\tau_{0}+ \epsilon \tau_{1}
\end{equation}
We expand the scalar field equation \eqref{scl} up to $\mathcal{O}(\epsilon)$ and obtain the coupled equations
\begin{equation}
    (r-r_H)^2\phi_1''+(r-r_H)\phi_1'=\frac{4\chi_0^2}{r^2}\Big[\phi_1 \alpha+ \chi_{1}\beta\Big],
\end{equation}
\begin{equation}
    (r-r_H)^2\chi_1''+(r-r_H)\chi_1'=\frac{4\chi_0^2}{r^2}\Big[\phi_1 \beta+ \chi_{1}\gamma\Big].
\end{equation}
where from \eqref{hessian}
\begin{align}\label{alphagammadef}
    &\alpha= \frac{\partial^2 V}{\partial \phi^2} \Big{|}_{\phi_{0}, \chi_{0}}=\frac{p^4}{8\sqrt{q^2p^2-(q.p)^2}}>0,\nonumber\\
    &\gamma=\,\frac{\partial^2 V}{\partial \chi^2} \Big{|}_{\phi_{0}, \chi_{0}}=\frac{p^4}{8\sqrt{q^2p^2-(q.p)^2}}>0\nonumber\\
    &\beta=\frac{\partial ^2 V}{\partial \phi \partial \chi}\Big{|}_{\phi_{0}, \chi_{0}}=0,
\end{align}
Since $\beta=0$ the equations for $\phi_1$ and $\chi_1$ decouples from each other. Solving these equations, we obtain
\begin{align}
\phi_{1}&=c_+\Big(\frac{r-r_{H}}{r}\Big)^{\delta_+}+c_-\Big(\frac{r-r_{H}}{r}\Big)^{\delta_-}\;, \nonumber \\
\chi_{1}&=d_+\Big(\frac{r-r_{H}}{r}\Big)^{\sigma_+}+d_-\Big(\frac{r-r_{H}}{r}\Big)^{\sigma_-}
\end{align}
where 
\begin{align}
    \delta_{\pm}&=\frac{1}{2}\Big(\pm \sqrt{1+\frac{16\chi_0^2\alpha}{r_H^2}}-1\Big)\;, \nonumber \\
    \sigma_{\pm}&=\frac{1}{2}\Big(\pm \sqrt{1+\frac{16\chi_0^2\gamma}{r_H^2}}-1\Big)
\end{align}
Demanding regularity of the scalar fields at the horizon, we need to set $c_{-}=0=d_{-}$. Putting the values of $\chi_{0}, \alpha$, $\gamma$ and $r_{H}$ from (\ref{phichisol}, \ref{alphagammadef}, \ref{rhdef}), we obtain
\begin{align}
    \delta_+=1=\sigma_+
\end{align}
 At asymptotic infinity $r\rightarrow \infty$, $\phi_1 \rightarrow c_+$ and $\chi_1\rightarrow d_+$. At $r\rightarrow r_{H}$, $\phi_1$ and $\chi_1$ becomes zero and hence $(\phi,\chi) \rightarrow (\phi_{0}, \chi_{0})$ irrespective of the value of $c_+$ and $d_+$. This is the attractor mechanism that works in the first-order perturbation theory.

 Let us now look at the corrections to the metric coefficients, which we expand to $\mathcal{O}(\epsilon)$ as follows:
 \begin{align}
    a=\,&a_0+\epsilon\, a_{1}\,,\nonumber\\
    b=\,&b_0+\epsilon \,b_1\,.
\end{align}
From \eqref{bequation}, one can easily see that $(b_1)^{\prime\prime}=0$, which implies that up to $\mathcal{O}(\epsilon)$, the metric coefficient $b$ can be written as $b=r+\epsilon(A_1r+A_2)$. Using the r-reparametrization symmetry, one can choose $A_1=A_2=0$, which is equivalent to taking $b_1=0$. The equation \eqref{mm} can then be used to solve for $a_1$ as $a_1=\frac{C_1}{r-r_H}+\frac{C_2}{r(r-r_H)}$. Demanding regularity at the horizon, forces us to set $C_1=C_2=0$. Thus we see that r-reparametrization and regularity at the horizon means that the metric coefficients does not receive any corrections at $\mathcal{O}(\epsilon)$ and starts receiving corrections at $\mathcal{O}(\epsilon^2)$. Let us calculate these corrections by writing 
\begin{align}
    a=\,&a_0+\epsilon^2 a_2\,,\nonumber\\
    b=\,&b_0+\epsilon^2 b_2\,.
\end{align}
Expanding \eqref{mm} upto order $\mathcal{O}(\epsilon^2)$, and demanding regularity at the horizon, we get
\begin{equation}\label{a2eq}
    a_2= -(1-\frac{r_{H}}{r})\frac{b_2}{r}
\end{equation}
Expanding Eq. \eqref{bequation} upto order  $\mathcal{O}(\epsilon^2)$, we get
\begin{equation}
    {b_2}''= -r G_{\tau\Bar{\tau}}(\phi_{0},\chi_{0})(\phi_{1}'^2+ \chi_{1}'^2)
\end{equation}
Using the already found solution for $\phi_1$ and $\chi_1$ on the R.H.S, the most general solution for $b_2$ is obtained as:
\begin{align}
    b_2=\,& -\frac{(c_+^2+d_+^2) r}{8\chi_0^2}\Big(1-\frac{r_{H}}{r}\Big)^{2}+D_1 r+D_2
\end{align}
Again using r-reparametrization symmetry, we set $D_1=D_2=0$. This solution of $b_2$ can be used in \eqref{a2eq} to solve for $a_2$ and we obtain:
\begin{align}
b_2=\,&-\frac{1}{8\chi_{0}^2}(c_+^2+d_+^2)r\Big(1-\frac{r_{H}}{r}\Big)^2 \\
a_2=\,&-\frac{1}{8\chi_{0}^2}(c_+^2+d_+^2)\Big(1-\frac{r_{H}}{r}\Big)^3
\end{align}
The second-order corrections to the metric vanish at $r=r_H$, indicating that these are the family of solutions having the same horizon radius and Bekenstein-Hawking entropy.

We now demonstrate that the attractor mechanism persists to all orders in perturbation theory. To this end, we expand the fields as power series in the perturbation parameter $\epsilon$:
\begin{subequations}
  \begin{align}
\phi(r) &= \sum_{l=0}^{\infty} \epsilon^{l}\,\phi_{l}(r)\,, \\
\chi(r) &= \sum_{l=0}^{\infty} \epsilon^{l}\,\chi_{l}(r)\,, \\
a(r) &= \sum_{l=0}^{\infty} \epsilon^{l}\,a_{l}(r)\,, \\
b(r) &= \sum_{l=0}^{\infty} \epsilon^{l}\,b_{l}(r)\,.\label{bseries}
\end{align}  
\end{subequations}
Motivated by the first sub-leading terms in the solution, we assume the following ansatz for the $l$-th order corrections\footnote{The linear term in $r$ in the ansatz for $b_l$ is motivated by the fact that $b\to r$ as $r\to\infty$ due to asymptotic flatness since $b$ appears in the metric as $b^2d\Omega^2$ \eqref{metrican}.}:
\begin{subequations}\label{lansatz}
   \begin{align}
\phi_{l}(r) &= c_{l}\left(1 - \frac{r_H}{r}\right)^{l}\,, \\
\chi_{l}(r) &= d_{l}\left(1 - \frac{r_H}{r}\right)^{l}\,, \\
b_{l}(r) &= h_{l}\, r \left(1 - \frac{r_H}{r}\right)^{l}\,, \\
a_{l}(r) &= f_{l}\left(1 - \frac{r_H}{r}\right)^{l+1}\,,
\end{align} 
\end{subequations}
where $c_l$, $d_l$, $h_l$, and $f_l$ are constants to be determined order by order from the equations of motion in terms of the constants $c_+$ and $d_+$ as well the charges carried by the black hole.
 Note that, from the leading and the sub-leading solutions obtained before, we have the lower order coefficients as
\begin{align}\label{LOC}
    &c_{0}=\chi_{0}\;, ~ d_{0}=\phi_{0}\;, c_1=c_+\;, d_1=d_+\nonumber \\
    &h_{0}=1\;,   f_{0}=1\;, h_1=f_1=0\;, \nonumber \\
    &h_2=f_2=-\frac{1}{8\chi_{0}^2}(c_+^2+d_+^2)
\end{align}
For convenience, we list down the exact equations of motion below:
\begin{subequations}\label{eomex}
\begin{align}
  \frac{b''}{b}=\,&- \frac{1}{4\chi^2}(\phi'^2+\chi'^2)\label{eomb}\\
    (a^2b^2)''=\,&2\label{eoma}\\
(a^2b^2\phi')'-\frac{2\phi'\chi'a^2b^2}{\chi}=\,&\frac{4 \chi^2}{b^2}\frac{\partial V}{\partial \phi}\label{scl1}\,\\
(a^2b^2\chi')'+\frac{(\phi'^2-\chi'^2)a^2b^2}{\chi}=\,&\frac{4 \chi^2}{b^2}\frac{\partial V}{\partial \chi}\label{scl2}
\end{align}
\end{subequations}
We now substitute the ansatz \eqref{lansatz} into \eqref{eomex} and obtain a recursion relation for obtaining the higher order coefficients from the lower order coefficients \eqref{LOC} as follows:
%collect terms up to order $\mathcal{O}(\epsilon^n)$. Equating the coefficients at this order, we obtain a recursive relation for $h_n$ in terms of the lower-order coefficients $h_{l<n}$, $c_{l<n}$, and $d_{l<n}$. The result is
\begin{align}\label{rec}
    h_{n} &=-\frac{T_{n} - \tilde{T}_{n}}{4 \chi_{0}^{2}\, n(n-1)}\;, \nonumber \\
    f_{n}&=-\frac{1}{2}\sum_{l_1+l_2+s_1+s_2=n}f_{l_1}f_{l_2}h_{s_{1}}h_{s_{2}}\;, \nonumber \\
    c_{n}&=\frac{\mathcal{C}_{n}-\tilde{\mathcal{C}}_{n}}{\chi_{0}\Big[r_{H}^2n(n+1)-4\alpha \chi_{0}^2\Big]}\;, \nonumber \\
    d_{n}&=\frac{\mathcal{B}_{n}-\tilde{\mathcal{B}}_{n}}{\chi_{0}\Big[r_{H}^2n(n+1)-4\alpha \chi_{0}^2\Big]}
\end{align}
where\;, 
\begin{align}\label{recdef}
    T_n=\,&\sum_{\substack{l_1+l_2+s_1=n\\s_1>1}}4h_{s_1}s_1(s_1-1)d_{l_1}d_{l_2}\;,\nonumber \\
    \tilde{T}_{n}=\,&\sum_{\substack{l_1+l_2+s_1=n\\l_1,l_2\geq 1}}h_{s_{1}}(c_{l_{1}}c_{l_{2}}+d_{l_{1}}d_{l_{2}})\;, \nonumber \\
    \mathcal{C}_{n}=\,&4\sum_{k=2}^{n}\sum_{\substack{s_1+s_2+s_3\nonumber \\
    +l_1+\cdots +l_{k}=n\\  l_1, \cdots,l_{k}\geq1}
  }\frac{1}{k!}V^{(k+1)}(\phi_{0}, \chi_{0})  d_{s_1}d_{s_{2}}d_{s_{3}}c_{l_1}\cdots c_{l_{k}}\;, \nonumber \\
    \tilde{\mathcal{C}}_{n}=\,&\sum_{\sum_{i}^{8}l_{i}=n}\Big[r_{H}^2l_{8}(n+1)-2l_{7}l_{8}\Big] f_{l_1}f_{l_2}h_{l_{3}}h_{l_{4}}h_{l_{5}}h_{l_{5}}h_{l_{6}}d_{l_{7}}c_{l_{8}}\;, \nonumber \\
    \mathcal{B}_{n}=\,&4\sum_{k=2}^{n}\sum_{\substack{s_1+s_2+s_3 \nonumber \\
    +l_1+\cdots +l_{k}=n\\  l_1, \cdots,l_{k}\geq1}
  }\frac{1}{k!}V^{(k+1)}(\phi_{0}, \chi_{0})  d_{s_1}d_{s_{2}}d_{s_{3}}d_{l_1}\cdots d_{l_{k}}\;, \nonumber \\
    \tilde{\mathcal{B}}_{n}=\,&\sum_{\sum_{i}^{8}l_{i}=n}\Big[r_{H}^2(n+1)l_{8}d_{l_{7}}d_{l_{8}}+l_{7}l_{8}(c_{l_{7}}c_{l_{8}}-d_{l_{7}}d_{l_{8}})\Big] f_{l_1}f_{l_2}h_{l_{3}}h_{l_{4}}h_{l_{5}}h_{l_{5}}h_{l_{6}}
\end{align}

%\begin{align}
    %4\sum_{m=0}^{3} \binom{3}{m}\chi_{0}^{3-m}\,&\Bigg(\sum_{k\geq1}^{n-m}\frac{1}{k!}V^{k+1}(\phi_{0}, \chi_{0})\nonumber\\\,&\sum_{\substack{
%l_1+\cdots+l_k \\
%+\, s_1+\cdots+s_m = n
%}}c_{l_1}\cdots c_{l_{k}}d_{s_1}\cdots d_{s_{m}}\Bigg)
   % \end{align}
%Extracting the coefficient $c_{n}$
%\begin{align}
   % &\Big(1-\frac{r_H}{r}\Big)^n \Bigg[
       % c_n n(n+1)\chi_0 r_H^2 
        %+ \Bigg(\sum_{l+j=n}d_l 
       % + \sum_{l+j=n}e_l 
       % \nonumber\\
       % &+ \sum_{l+s+j=n}e_l d_s  + \sum_{l+m=n}e_l e_m \nonumber\\
       % & 
       % + \sum_{l+m+s=n}e_l e_m d_s\Bigg)
       % c_j j(j+1)r_H^2\nonumber\\
      %  & +\Bigg(2\sum_{l+m=n}c_ld_m+4\sum_{s+l+m=n}e_sc_ld_m \nonumber\\
      %  & +2\sum_{l+m+k+s}e_ke_sc_ld_m\Bigg)lm
  %  \Bigg].
%\end{align}
%and expanding the R.H.S of \eqref{scl1} upto order $\mathcal{O}(\epsilon^n)$
%\begin{align}
   % &4\Big(1-\frac{r_H}{r}\Big)^n\Bigg[c_n\chi_{0}^3 V^{(2)}+ \frac{1}{k!}\sum_{k=2}^{n}\sum_{l_1+\dots+l_k=n}c_{l_1}\dots c_{l_k}V^{(k+1)}\nonumber\\
   % &\qquad+ 3\chi_0 \frac{1}{k!}\sum_{k=1}^{n-2}\sum_{s_1+s_2+l_1+\dots+l_{k}=n}c_{l_1}\dots c_{l_k}d_{s_1}d_{s_2}V^{(k+1)}\nonumber\\
   % &\qquad +3\chi_0^2 \frac{1}{k!}\sum_{k=1}^{n-1}\sum_{s_1+l_1+\dots+l_{k}=n}c_{l_1}\dots c_{l_k}d_{s_1}V^{(k+1)}\nonumber\\
    %&\qquad +\frac{1}{k!}\sum_{k=1}^{n-3}\sum_{s_1+s_2+s_3+l_1+\dots+l_k=n}c_{l_1}\dots c_{l_{k}}d_{s_1}d_{s_2}d_{s_3}V^{(k+1)}
      %  \Bigg]
%\end{align}
In the above expressions, we have defined 
\begin{equation}
    V^{k+1}\equiv{\frac{\partial^{k+1} V}{\partial \phi^{k+1}}}\vert_{(\phi,\chi)=(\phi_0, \chi_0)}
\end{equation}
 We obtained all the coefficients of the power series expansion in terms of $c_+$ and $d_+$. All the corrections to all the fields vanish at $r=r_H$. Hence, the horizon works as an attractor for all order perturbative corrections. Near the horizon, $r>r_{H}$, we need to check the convergence of the series. We do this by the method of induction (See Appendix-\ref{convergence}). The solutions can be seen as a power series expansion in the nearness parameter $\xi=\frac{(r-r_H)}{r}$. Near the horizon, the dominant contribution comes from the first-order corrections. In the following section,
we use these first-order corrections to fix the boundary conditions near the horizon and numerically solve the exact field equations.
\section{Numerical analysis}
For numerical analysis, we choose the charge configuration as follows
\begin{equation*}
p=(1,2,0,0,0,0),~~~q=(3,1,0,0,0,0)
\end{equation*}
The horizon radius is 
\begin{equation*}
  r_H=2(p^2q^2-(p.q)^2)^{1/4}=\sqrt{20}=4.472
\end{equation*}
The constant moduli solution is completely fixed by the charge configuration, where the axion-dilaton moduli takes the following values
\begin{equation*}
    \phi_{0}=\frac{4 \,q.p}{p^2}=-4,~~\chi_{0}=-\frac{4\sqrt{p^2q^2-(p.q)^2}}{p^2}=4
\end{equation*}
Now we wish to solve the equations exactly for $r\geq r_H$ in general for different asymptotic boundary conditions of the moduli fields by integrating the equations numerically. However, it is not possible to give asymptotic boundary conditions numerically. So we follow the approach of \cite{Goldstein:2005hq} and apply boundary conditions near the horizon given below, which we obtain using the perturbative solutions obtained in the previous section up to first subleading order.
\begin{subequations}
    \begin{align}
        \phi\Big(\frac{r_{H}}{1-\xi}\Big)=\,&\phi_{0}+c_+\xi\\
        \chi\Big(\frac{r_{H}}{1-\xi}\Big)=\,&\chi_{0}+d_+\xi\\
        a\Big(\frac{r_{H}}{1-\xi}\Big)=\,& \xi-\frac{1}{8\chi_{0}^2}(c_+^2+d_+^2)\xi^3\\
        b\Big(\frac{r_{H}}{1-\xi}\Big)=\,&\frac{r_{H}}{1-\xi}-\frac{1}{8\chi_{0}^2}(c_+^2+d_+^2)r_{H}\frac{\xi^2}{1-\xi}
    \end{align}
\end{subequations}
where, $\xi=\frac{r-r_H}{r}$ is the nearness parameter. Using these boundary conditions, we numerically integrate the exact equations \eqref{scl1}, \eqref{scl2}, \eqref{bequation} and \eqref{mm} and obtain the following plots which demonstrate the attractor behaviour of the axion-dilaton moduli.
\begin{figure}[H]
\centering % Move \centering above the \includegraphics for clarity
\includegraphics[width=8.5cm]{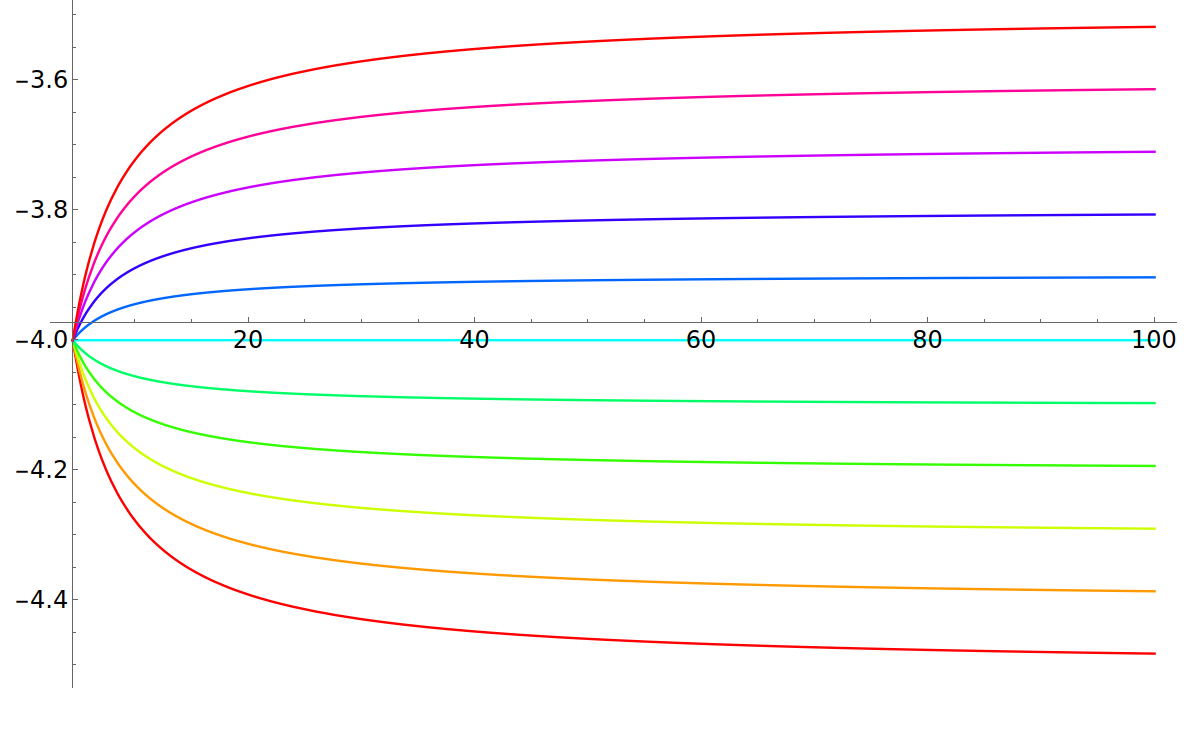}
\caption{The attractor flow of $\phi(r)$ with attractor value $-4$ in the range $r\in [4.472, 50]$. The nearness parameter $\xi=2\times 10^{-8}$,   $d_+=0.05$ and the different colors corresponds to different $c_+=[-0.5,0.5]$ with stepsize $0.1$.}
\label{fig:axion1} % Optional, for referencing
\end{figure}
\begin{figure}[H]
\centering % Move \centering above the \includegraphics for clarity
\includegraphics[width=8.5cm]{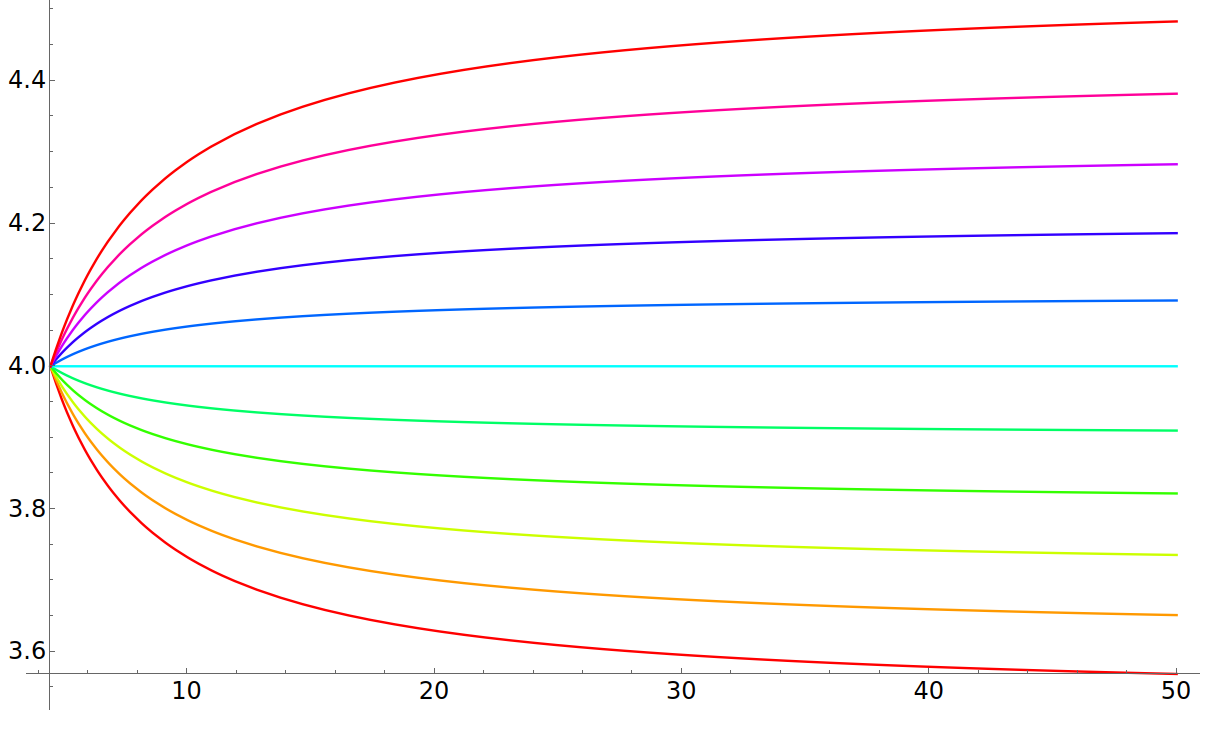}
\caption{The attractor flow of $\chi(r)$ with attractor value $4$ in the range $r\in [4.472, 50]$. The nearness parameter $\xi=2\times 10^{-8}$,   $c_+=0$ and the different colours corresponds to different $d_+=[-0.5,0.5]$ with stepsize $0.1$.}
\label{fig:dilaton1} % Optional, for referencing
\end{figure}
\begin{figure}[H]
\centering % Move \centering above the \includegraphics for clarity
\includegraphics[width=8.5cm]{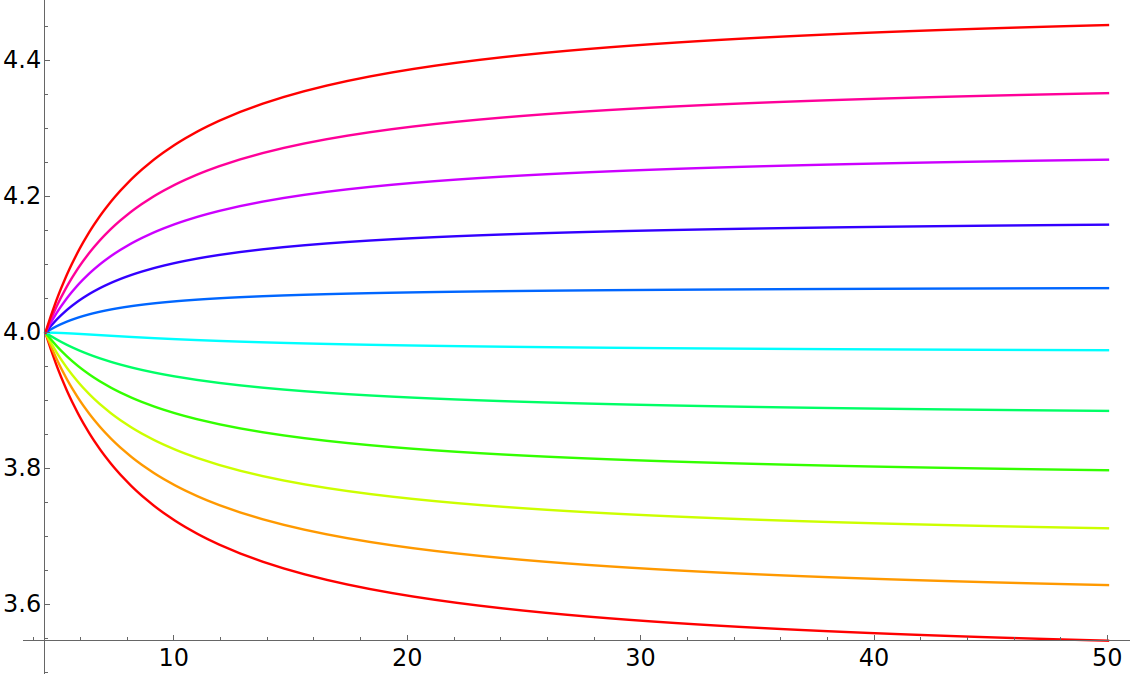}
\caption{The attractor flow of $\chi(r)$ with attractor value $4$ in the range $r\in [4.472, 50]$. The nearness parameter $\xi=2\times 10^{-8}$,   $c_+=0.5$ and the different colors corresponds to different $d_+=[-0.5,0.5]$ with stepsize $0.1$.} %The blue line corresponds to $d_1=0$ which is not a constant solution unlike Figure \ref{fig:dilaton1}. This is due to the presence of the non-linear term $(\phi')^2$ in \eqref{scl2} which goes to zero near the horizon for $c_1=0$ and allows a constant solution for $\chi(r) $. However in \eqref{scl1}, no such term survives near the horizon.} 
\label{fig:ghj}
\end{figure}
\section{Supersymmetry property of the constant moduli solution}\label{susyconstant}
In this section, we analyse the supersymmetry of the constant moduli solutions in pure $\NN=4$ supergravity where all the six electric and magnetic charges are turned on. We perform the analysis using the superconformal formalism. In order to go from conformal to Poincar{\'e} supergravity, one needs to gauge fix the additional symmetries such as dilatation, special conformal transformations, $SU(4)$ R-symmetry as well as S-supersymmetry. We impose explicit gauge fixing conditions for the first three and work with the modified supersymmetry transformation rules. However, as far as $S$-supersymmetry is concerned, we do not choose an explicit gauge fixing condition. Rather, we use an appropriate $S$-supersymmetry compensator and construct combination of fermionic fields that are invariant under $S$-supersymmetry. This approach has been followed in \cite{LopesCardoso:2000qm} and recently in \cite{Bhattacharjee:2025qro, Adhikari:2025eok} to find supersymmetric solutions in higher derivative supergravity theories. 

To construct the $S$-invariant fermions we need to find an appropriate compensating fermionic field. Such a fermion for $\mathcal{N}=4$ supergravity has been constructed in \cite{Bhattacharjee:2025qro} and is given as:
\begin{equation}\label{in}
    \zeta_{i}= 2\varphi^{-2} \phi_{ij}^{I}{\psi^{j J}}\eta_{IJ},
\end{equation}
where 
\begin{align}
    \varphi^2=\phi^{Iij}\phi^{J}_{ij}\eta_{IJ}
\end{align}
The $Q$ and $S$ -supersymmetry transformation of this fermion are given as 
\begin{align}\label{compensator}
    \delta \zeta^{i}= \,&2\varphi^{-2}\phi^{I ij}\Big[-\frac{1}{2\Phi}\gamma^{ab}\epsilon_{j}\left({F}_{ab}^{+J}+\Phi^{*}T_{ablk}\phi^{Jlk}\right)\nonumber\\ \,&-2\slashed{D}\phi^{J}_{jk}\epsilon^{k} + \epsilon_{k}E_{jl}\phi^{Jlk}\Big]\eta_{IJ}+\eta^{i}.
\end{align}
Now we write down the constant moduli solution in a form that are convenient for the supersymmetry analysis. The index convention is as follows: $(\mu, \nu)$ are four-dimensional world indices, and $\underline{\mu}, \underline{\nu}$ are three-dimensional world indices. ($a,b,..$) are four dimensional Lorentz indices and ($\underline{a}, \underline{b},..$) are three dimensional Lorentz indices. It is convenient to write the metric in the IWP form:
\begin{align}
    ds^2=-e^{2g}(dt+\omega)^2+e^{-2g}dx_{\underline{\mu}}dx^{\underline{\mu}}
\end{align}
We recall that the metric is given as
\begin{align}
   ds^2=-a(r)^2dt^2+ a(r)^{-2}dr^2+b(r)^2d\Omega^2
\end{align}
where 
\begin{align}
    a(r)=1-\frac{r_{H}}{r},~~~ b(r)=r,~~~ r_{H}^2=4\sqrt{p^2q^2-(p.q)^2} 
\end{align}
With the following coordinate transformation
\begin{align}
    r=\rho+r_{H}
\end{align}
the metric takes the IWP form
\begin{align}
    ds^2=-\Big(\frac{\rho}{\rho+r_{H}}\Big)^2dt^2+\Big(1+\frac{r_{H}}{\rho}\Big)^2(d\rho^2+\rho^2 d\Omega^2)
\end{align}
with the following identification
\begin{align}
    e^{-g}=\Big(1+\frac{r_{H}}{\rho}\Big), ~~\omega=0
\end{align}
The components of the inverse vierbein are given by
\begin{align}
    e_{0}{}^{t}=e^{-g},~~e_{0}{}^{\underline{\mu}}=0,~~e_{\underline{a}}{}^{t}=0,~~e_{\underline{a}}{}^{\underline{\mu}}=e^{g}\hat{e}_{\underline{a}}{}^{\underline{\mu}}
\end{align}
The rescaled dreibein components $\hat{e}_{\underline{a}}{}^{\underline{\mu}}$ of the three dimnensional flat base space are given by
\begin{align}
    \hat{e}_{\rho}{}^{1}=1,~~\hat{e}_{\theta}{}^{2}=\rho,~~\hat{e}_{\phi}{}^{3}=\rho\sin{\theta}
\end{align}
The non-zero spin connection coefficients are 
\begin{subequations}
    \begin{align}
    \omega_{\underline{a}\,\underline{b}\,\underline{c}}
    \equiv\,& e_{\underline{a}}{}^{\mu}\omega_{\mu bc}=\,e^{g}\Big(\hat{\omega}_{\underline{a}\,\underline{b}\,\underline{c}}+2\delta_{\underline{a}[\underline{b}}\hat{\nabla}_{\underline{c}]}g\Big)\\
    \omega_{00\underline{a}}\equiv\,&e_{0}{}^{\mu}\omega_{\mu 0a}=e^{g}\hat{\nabla}_{\underline{a}}g
\end{align}
\end{subequations}
where $\hat{\omega}_{\underline{a}\,\underline{b}\,\underline{c}}\equiv \hat{e}_{\underline{a}}{}^{\underline{\mu}}\hat{\omega}_{\underline{\mu}\underline{b}\underline{c}}$ is the spin connection coefficient associated with the three dimensional base space:
%\begin{align}
  %  \hat{\omega}_{\underline{\mu}}{}^{pq}=\hat{e}^{\underline{\nu}~p}\hat{e}^{\underline{\sigma}~q}\hat{e}_{\underline{\mu}}{}^{l}\partial_{[\underline{\nu}}\hat{e}_{\underline{\mu}]l}-\hat{e}^{\underline{\nu}~p}\partial_{[\underline{\nu}}\hat{e}_{\underline{\mu}]}{}^{q}+\hat{e}^{\underline{\nu}~q}\partial_{[\underline{\nu}}\hat{e}_{\underline{\mu}]}{}^{p}
%\end{align}
The non-zero components are given by
\begin{align}
    \hat{\omega}_{\theta12}=\,&1\\
    \hat{\omega}_{\phi13}=\,&\sin{\theta}\\
    \hat{\omega}_{\phi23}=\,&\cos{\theta}
\end{align}
The covariant derivative $\hat{\nabla}_{\underline{a}}=\hat{e}_{\underline{a}}{}^{\underline{\mu}}\hat{\nabla}_{\underline{\mu}}$ is covariant w.r.t the local rotations in the three-dimensional base space.
% \begin{align}
%     \nabla_{\underline{\mu}}=\partial_{\underline{\mu}}-\frac12 \hat{\omega}_{\underline{\mu}}{}^{\underline{a}\underline{b}}M_{\underline{a}\underline{b}}=\hat{e}_{\underline{\mu}}{}^{\underline{a}}\nabla_{\underline{a}}
% \end{align}

The field strength components in the constant moduli solutions are given as:
\begin{subequations}\label{Fieldstrconmod}
    \begin{align}
{F}_{t\rho}^{I}=\,&
e^{I}=\frac{f^{IJ}(q_{J}+i\Tilde{f}_{JK}p^K)}{(\rho+r_{H})^2}\nonumber\\\,&=\frac{1}{(\rho+r_{
H})^2\chi_{0}}\Big(-4\eta^{IJ}q_{J}+\phi_{0} p^{I}\Big)\\
{F}_{\theta\phi}^{I}=\,& p^{I}\sin{\theta}\\
{F}_{t\rho}^{I\mp}=\,&\frac{1}{2(\rho+r_{H})^2}\Big(\frac{-4\eta^{IJ}q_{J}+\phi_{0}p^{I}}{\chi_{0}}\pm i~p^{I}\Big)
\end{align}
\end{subequations}
and the axion-dilaton moduli are given by
\begin{equation}
    \phi_0=4\frac{q.p}{p^2},~~\chi_{0}=-4\frac{\sqrt{q^2p^2-(p.q)^2}}{p^2}
\end{equation}
Since the moduli fields are constant throughout spacetime, Eqs. \eqref{pmutau} and \eqref{taumu} imply
\begin{align}\label{pmuamu=0}
P_{\mu} = 0 = a_{\mu}.
\end{align}
For pure $\NN=4$ supergravity, the scalar fields $\phi^{Iij}$ are constants and satisfies \eqref{purephi}. The auxiliary field equations given in \eqref{vmuij} and \eqref{eomdijkl} implies
\begin{subequations}
    \begin{align}
        V_{\mu}{}^{i}{}_{j}=\,&0\label{vmupure}\\
        D^{ij}{}_{kl}=\,&0\label{dijklpure}
    \end{align}
\end{subequations}
We now substitute the values of some of the auxiliary fields from \eqref{eij}, \eqref{vmupure} and \eqref{dijklpure} together with \eqref{pmuamu=0} into the fermionic transformation rules given in \eqref{trans} and obtain
\begin{subequations}
    \begin{align}
     \delta \psi_\mu{}^i=\,&2\,\mathscr{D}_\mu\epsilon^i
   -\tfrac12\gamma^{ab}T_{ab}{\!}^{ij}\gamma_\mu\epsilon_j-\gamma_\mu\eta^i\\
   \delta \chi^{ij}{}_{k}=\,&-\tfrac12\gamma^{ab}
   \slashed{D}T_{ab}{\!}^{ij}\epsilon_k\\
   \delta \Lambda_{i}=\,&\frac{1}{2}\varepsilon_{ijkl}T_{bc}{}^{kl}\gamma^{bc}\epsilon^{j}
\end{align}
\end{subequations}
We will substitute the value of the auxiliary field $T_{ab}{}^{ij}$ from the field equation \eqref{auxitab} later, in our analysis.

Next, we impose the gauge fixing conditions on the transformation rules. We begin with the special conformal transformation (SCT) gauge condition $b_{\mu}=0$. As a consequence of this and (\ref{vmupure}, \ref{pmuamu=0}), equation \eqref{cdepsilon} only contains the standard spin connection
\begin{align}
    \mathscr{D}_{\mu}\epsilon^{i}=\,&\Big(\partial_{\mu}-\frac{1}{4}\omega_{\mu}{}^{ab}\gamma_{ab}\Big)\epsilon^{i}
\end{align}
Since the $S$-supersymmetry transformation of the gravitino is non-zero, we need to construct the $S$-invariant combination $\psi_{\mu}{}^{i}+\gamma_{\mu}\zeta^{i}$. However the $Q$- supersymmetry transformation of $\zeta^{i}$ \eqref{compensator} vanishes upon imposing the auxiliary field equations and one is left only with the $S$-supersymmetry: 
\begin{align}
    \delta \zeta^{i}=\eta^{i}
\end{align}
Hence the analysis does not require us to explicitly construct $S$-invariant spinors by adding terms proportional to $\zeta^i$.\footnote{This will however not be the case for matter coupled supergravity since the $Q$-supersymmetry transformation of $\zeta^i$ will not vanish in that case.} The $Q$- supersymmetry transformation of the gaugino ($\psi^{I}_i$) belonging to the vector multiplet also vanishes upon using the auxiliary field equations. Hence, the non-trivial Killing spinor equations are obtained from setting the $Q$-supersymmetry variations of the Weyl multiplet fermions ($\psi_{\mu}^{i},\Lambda_i,\chi^{ij}{}_{k}$) to zero. They read
\begin{align}
    \Big(\partial_{\mu}-\frac{1}{4}\omega_{\mu}{}^{ab}\gamma_{ab}\Big)\epsilon^{i}  -\tfrac12\gamma^{ab}T_{ab}{\!}^{ij}\gamma_\mu\epsilon_j=\,&0\label{Killing1}\\
    \gamma^{ab}
   \slashed{D}T_{ab}{\!}^{ij}\epsilon_k=\,&0\label{Killing2}\\
   \varepsilon_{ijkl}T_{bc}{}^{kl}\gamma^{bc}\epsilon^{j}=\,&0\label{Killing3}
\end{align}
Now we analyse \eqref{Killing1} component-wise. The time-component reads
\begin{align}
2\partial_{t}\epsilon^{i}-\frac{1}{2}\omega_{t}{}^{ab}\gamma_{ab}\epsilon^{i}-\frac{1}{2}\gamma^{ab}T_{ab}{}^{ij}\gamma_{t}\epsilon_{j}=0
\end{align}
Since we are considering static solutions, the Killing spinors are independent of time. Using the explicit form of the spin connection coefficients, we obtain
\begin{align}\label{con1}
    T_{\underline{a}0}{}^{ij}\epsilon_{j}=-\frac{1}{2}e^{g}\hat{\nabla}_{\underline{a}}g \gamma_{0}\epsilon^{i}
\end{align}
From the space-component of \eqref{Killing1} 
\begin{align}\label{spacecom}
   2\partial_{\underline{\mu}}\epsilon^{i}-\frac{1}{2}\omega_{\underline{\mu}}{}^{ab}\gamma_{ab}\epsilon^{i}-\frac{1}{2}\gamma^{ab}T_{ab}{}^{ij}\gamma_{\underline{\mu}}\epsilon_{j}=0 
\end{align}
We expand the second and third terms as follows
\begin{subequations}
    \begin{align}
    -\frac{1}{2}\omega_{\underline{\mu}}{}^{ab}\gamma_{ab}\epsilon^{i}=\,&-\frac{1}{2}e^{g}{e}_{\underline{\mu}}{}^{\underline{a}}\hat{\omega}_{\underline{a}}{}^{\underline{b}\,\underline{c}}\gamma_{\underline{b}\,\underline{c}}\epsilon^{i}\nonumber\\\,&-e_{\underline{\mu}}{}^{\underline{a}}e^{g}\hat{\nabla}^{\underline{b}}g \gamma_{\underline{a}\,\underline{b}}\epsilon^{i}\\
     - \frac{1}{2}\gamma^{ab}T_{ab}{}^{ij}\gamma_{\underline{\mu}}\epsilon_{j}=\,&2~e_{\underline{\mu}}{}^{\underline{a}}T_{\underline{a}0}{}^{ij}\gamma^{0}\epsilon_{j}\nonumber\\\,&+2~e_{\underline{\mu}}{}^{\underline{a}}T_{\underline{a}\,\underline{b}}{}^{ij}\gamma^{\underline{b}}\epsilon_{j}
    \end{align}
\end{subequations}
Substituting \eqref{con1} in \eqref{spacecom} yields
\begin{align}
  2\partial_{\underline{\mu}}\epsilon^{i}-\frac{1}{2}e^{g}{e}_{\underline{\mu}}{}^{\underline{a}}\hat{\omega}_{\underline{a}}{}^{\underline{b}\,\underline{c}}\gamma_{\underline{b}\,\underline{c}}\epsilon^{i}-\partial_{\underline{\mu}}g\,\epsilon^{i}=0
\end{align}
This yields
\begin{subequations}
    \begin{align}
        2\partial_{\theta}\epsilon^{i}-\gamma_{12}\epsilon^{i}=\,&0\label{equ1}\\
        2\partial_{\phi}\epsilon^{i}-\Big(\sin{\theta}\,\gamma_{13}+\cos{\theta}\,\gamma_{23}\Big)\epsilon^{i}=\,&0\label{equ2}\\
       2\partial_{\rho}\epsilon^{i}-(\partial_{\rho}g)~\epsilon^{i}=\,&0\label{equ3}
    \end{align}
\end{subequations}
The above equations can be solved by separation of variables and one obtains: 
% \begin{align}
%     \epsilon^{i}=e^{\frac{\theta}{2}\gamma_{12}}X^{i}(\rho, \phi)
% \end{align}
% Putting this back in \eqref{equ2} yields 
% \begin{align}
%     \partial_{\phi}X^{i}(\rho, \phi)-\frac{1}{2}e^{-\frac{\theta}{2}\gamma_{12}}(\sin{\theta}\gamma_{13}+\,&\cos{\theta}\gamma_{23})e^{\frac{\theta}{2}\gamma_{12}}X^{i}(\rho, \phi)\nonumber\\\,&=0
% \end{align}
% Now using algebra of gamma matrices, one can show that 
% \begin{align}
%     e^{-\frac{\theta}{2}\gamma_{12}}(\sin{\theta}\gamma_{13}+\cos{\theta}\gamma_{23})e^{\frac{\theta}{2}\gamma_{12}}=\gamma_{23}
% \end{align}
% which implies that 
% \begin{align}
%     X^{i}(\rho, \phi)=e^{\frac{\phi}{2}\gamma_{23}}Y^{i}(\rho)
% \end{align}
% Finally solving \eqref{equ3}, we obtain 
% \begin{align}\label{Killing}
%     Y^{i}(\rho)=e^{\frac{g}{2}}\epsilon_{(0)}^{i}=\Big(1+\frac{r_{H}}{\rho}\Big)^{-1/2}\epsilon_{(0)}^{i}
% \end{align}
% Then the full solution is 
\begin{align}
    \epsilon^{i}=e^{\frac{1}{2}\theta \gamma_{12}+\frac{1}{2}\phi \gamma_{23}}e^{g/2}\epsilon^{i}_{0}
\end{align}
where $\epsilon_{(0)}^{i}$ are constant spinors.
The angular dependence of the Killing spinor is a coordinate artefact that can be removed by a local rotation as $\epsilon^{i}\rightarrow U \epsilon^{i}$ where 
\begin{align}
    U(\theta,\phi)=e^{-\frac{1}{2}\theta \gamma_{12}-\frac{1}{2}\phi\gamma_{23}}
\end{align}
So we can write the final solution as
\begin{align}
    \epsilon^{i}=e^{g/2}\epsilon^{i}_{(0)}
\end{align}
The second Killing spinor equation \eqref{Killing2} does not give any additional constraint. It is identically satisfied upon putting the above solution for the Killing spinor because of the Maxwell's equation \eqref{Tabmaxwell}. 
%  Now we put the solution for $\epsilon^{i}$ into \eqref{Killing2}
% \begin{align}
%     \gamma_{d}D_{c}T^{cdij}\epsilon_{(0)k}=0
% \end{align}
% This equation is identically zero, since from \eqref{Tabmaxwell} we obtain that 
% \begin{align}
%     D_{c}T^{cd ij}=0
% \end{align}
% So this equation does not put any constraints on the parameters $\epsilon^{i}_{(0)}$.

Putting the auxiliary field equation for $T_{ab}{}^{ij}$ \eqref{auxitab} and the field strengths for the constant moduli solutions \eqref{Fieldstrconmod}, the third Killing spinor equation \eqref{Killing3} and the constraint \eqref{con1} can be written as
% \begin{align}
%    \hat{F}_{p0}^{I+}\phi^{Jij}\eta_{IJ}\epsilon_{(0)j}=\,&0\label{firstcon}
% \end{align}
% The above can be written as 
\begin{subequations}\label{ABeq}
    \begin{align}
     A^{ij}\epsilon_{(0)j}=\,&0\label{conkilliing}\\
      B^{ij}\epsilon_{(0)j}=\,&\frac{r_{H}}{\sqrt{\chi_{0}}} \gamma_{0}\epsilon_{(0)}^{i}\label{conkilliing1}
\end{align}
\end{subequations}
where 
\begin{subequations}\label{ABdef}
    \begin{align}
    A^{ij}\equiv\mathcal{Q}^{I}\phi^{K~ij}\eta_{IK}, ~~\mathcal{Q}^{I}= \Big(\frac{-4\eta^{IJ}q_{J}+\phi_{0}p^{I}}{\chi_{0}}-i~p^{I}\Big)\label{Amatrix}\\\
     B^{ij}\equiv\bar{\mathcal{Q}}^{I}\phi^{K~ij}\eta_{IK}, ~~\bar{\mathcal{Q}}^{I}= \Big(\frac{-4\eta^{IJ}q_{J}+\phi_{0}p^{I}}{\chi_{0}}+i~p^{I}\Big)\label{Bmatrix}
\end{align}
\end{subequations}

%Using the equations of motion and \eqref{Killing}, we get 
%\begin{align}
%\hat{F}_{p0}^{I+}\phi^{Jij}\eta_{IJ}\epsilon_{(0)j}=\,&0\label{firstcon}\\
%\hat{F}_{p0}^{I-}\phi^{Jij}\eta_{IJ}\epsilon_{(0)j}=\,& \frac{1}{2}\Phi e^{g}\nabla_{p}g \gamma_{0}\epsilon_{(0)}^{i}\label{secondcon}
%\end{align}
%So we can write the first condition as 
%\begin{align}
   % \Big(\frac{-\eta^{IJ}q_{J}+\phi_{0}p^{I}}{\chi_{0}}-i~p^{I}\Big)\phi^{K ~ij}\eta_{IK}\epsilon_{(0)j}=0\nonumber\\
  %\implies  A^{ij}\epsilon_{(0)j}=0\label{conkilliing}
%\end{align}
We can show that the anti-symmetric matrix $A^{ij}$ is singular using the fact that the determinant of anti-symmetric matrix is equal to the square of its Pfaffian ($Pf(A)$) where $Pf(A)$ is explicitly shown to vanish as follows:
\begin{align}\label{pfaffcond}
Pf(A)=\varepsilon_{ijkl}A^{ij}A^{kl}=\,&\varepsilon_{ijkl}\mathcal{Q}^{I}\mathcal{Q}^{J}\phi^{K~ij}\phi^{L~kl}\eta_{IK}\eta_{JL}\nonumber\\
=\,&\mathcal{Q}^{I}\mathcal{Q}^{J}\eta_{IJ}\nonumber \\
=\,&\frac{1}{\chi_{0}^2}\Big(16q^2-8(q.p)\phi_{0}+\phi_{0}^2 p^2\Big)-p^2
=0
\end{align}
where, in the last step we have used the constant moduli solutions for $\phi_0$ and $\chi_0$. Similarly, we can show that $Pf(B)$ also vanishes.

Hence, the anti-symmetric matrices $A^{ij}$ and $B^{ij}$ are of rank 2. They transform in the $\mathbf{6}$ representation of $SU(4)$ R-symmetry. In the next subsection, we will choose an appropriate $SU(4)$ R-symmetry gauge fixing condition on the scalars $\phi^{Iij}$. This will determine the matrices $A^{ij}$ and $B^{ij}$ completely in terms of the charges. 

% Since $\phi^{I}_{ij}$ transforms in the $\mathbf{6}$ representation of $SU(4)$, it follows that $A^{ij}$ also transforms in same representation. This implies that there exists an $SU(4)$ basis where $A^{ij}$ can be written in a block diagonal form, with one block equal to the zero matrix. In the next subsection, we find such a basis explicitly in terms of the black hole charges.
\subsection{$SU(4)$ gauge fixing}
It is convenient to go from the $SU(4)$ representation of $\phi_{ij}^{I}$ to the $SO(6)$ representation via the following:
\begin{align}
     \phi^{I}_{ij}=x_{A}^{I}\Gamma_{A}{}_{ij}
\end{align}
where $A=1,\cdots, 6$. $x_{A}^{I}$ transforms under the vector-representation of $SO(6)$ and the matrices $\Gamma_{A}$ are given as \cite{deRoo:1984zyh, Gliozzi:1976qd, Brink:1976bc}:
\begin{align}\label{rep}
   \Gamma_{A}= \begin{pmatrix}
        \beta_{A}, ~~\text{for} ~~1\leq A\leq 3 \\
        i\alpha_{A}, ~~\text{for} ~~4\leq A\leq 6
    \end{pmatrix}
\end{align}
where $\alpha_{A}$ and $\beta_{A}$ span an $SU(2)\times SU(2)$ algebra and are given as:
\begin{align}
    \beta_1=\begin{pmatrix}
        0& i\sigma^2\\
        i\sigma^2&0
    \end{pmatrix},~~\beta_2= \begin{pmatrix}
        0& 1\\
        -1&0
    \end{pmatrix},~~\beta_{3}=\begin{pmatrix}
        -i\sigma^2& 0\\
        0&i\sigma^2
    \end{pmatrix}
\end{align}
\begin{align}
    \alpha_1=\begin{pmatrix}
        0& \sigma^1\\
        -\sigma^1&0
    \end{pmatrix},~~\alpha_2= \begin{pmatrix}
        0& -\sigma^3\\
        \sigma^3&0
    \end{pmatrix},~~\alpha_{3}=\begin{pmatrix}
        i\sigma^2& 0\\
        0&i\sigma^2
    \end{pmatrix}
\end{align}
The $\Gamma$-matrices satisfy the following conditions 
\begin{subequations}
    \begin{align}
   (\Gamma_{Aij})^{*}\equiv\,&{\Gamma}^{ij}_{A}=-\frac{1}{2}\varepsilon^{ijkl}\Gamma_{Akl}\\
   {\Gamma}^{ij}_{A}\Gamma_{Bji}=\,&4\delta_{AB}\\
   {\Gamma}^{kl}_{A}\Gamma_{Aij}=\,&-4\delta^{k}_{[i}\delta^{l}_{j]}
\end{align}
\end{subequations}
Along with the above relations, the pseudo-reality condition on $\phi^{I}_{ij}$ (see-Table-\ref{tab:vector}) implies that the fields $x_{A}^{I}$ are real. As a consequence 

\begin{align}
\phi^{Iij}=x_{A}^{I}{\Gamma}_{A}^{ij}    
\end{align}
Equation-\eqref{dijkl}, encompassing the dilatation gauge fixing condition and the constraints imposed by the Lagrange multiplier $D^{ij}{}_{kl}$, written in terms of $\phi^{Iij}$, translates to the following equation in terms of $x_{A}^{I}$.
\begin{align}
    x_{A}^{I}\eta_{IJ}x_{B}^{J}=-\frac{1}{4}\delta_{AB}
\end{align}
The above equation is still invariant under the local $SU(4)\sim SO(6)$ transformation which we are yet to fix. For pure supergravity ($1\leq I\leq 6$), one can exploit the local $SO(6)$ symmetry to fix the fields $x_{A}^{I}$ to the following constants, 
\begin{align}
    x_{A}^{I}=\frac{1}{2}\delta_{A}^{I}
\end{align}
% With this we have
% \begin{align}
%      \phi^{I}_{ij}=\frac{1}{2}\Gamma^{I}_{ij}
% \end{align}
% and \begin{align}
%     \phi^{I~ij}=-\frac{1}{2}\tilde{\Gamma}^{Iij}=-\frac{1}{2}(\tilde{\Gamma}^{I}_{ij})^{*}
% \end{align}
% Next, decompose the $\Gamma$ matrices as follows
Upon using the above $SO(6)$ gauge fixing condition, the matrices $A$ and $B$ \eqref{ABdef} take the form  
\begin{align}\label{matrixA}
    A=\,&\begin{pmatrix}
        0&&-P^{4}&&P^2&& P^3\\
       P^{4}&&0&&-P^{6}&&P^{5}\\
        -P^2&&P^{6}&&0&& P^{1}\\
      -P^3&&-P^{5}&&-P^{1}&&0
    \end{pmatrix}\\
       B=\,&\begin{pmatrix}
        0&&-P^{1*}&&P^{5*}&& P^{6*}\\
      P^{1*}&&0&&-P^{3*}&&P^{2*}\\
        -P^{5*}&&P^{3*}&&0&& P^{4*}\\
      -P^{6*}&&-P^{2*}&&-P^{4*}&&0
    \end{pmatrix}
\end{align}
where we have defined
\begin{align}
    P^1\equiv\,&\frac{1}{2}(\mathcal{Q}^3-i\mathcal{Q}^6), ~P^4\equiv\frac{1}{2}(\mathcal{Q}^3+i\mathcal{Q}^6)\\
    P^2\equiv\,&\frac{1}{2}(\mathcal{Q}^2-i\mathcal{Q}^5), ~P^5\equiv\frac{1}{2}(\mathcal{Q}^2+i\mathcal{Q}^5)\\
    P^3\equiv\,&\frac{1}{2}(\mathcal{Q}^1-i\mathcal{Q}^4), ~P^6\equiv\frac{1}{2}(\mathcal{Q}^1+i\mathcal{Q}^4)
\end{align}
The Pfaffian condition \eqref{pfaffcond} reads
\begin{align}\label{Pfian}
    P^1P^4+P^2P^5+P^3P^6=0
\end{align}
Now, we observe that the equations \eqref{conkilliing} and \eqref{conkilliing1} are invariant under the following transformations
\begin{align}\label{transf}
  A^{ij}\rightarrow \,&\tilde{A}^{ij}=U^{i}{}_{k}A^{kl}(U^{T})_{l}{}^{j}\nonumber\\
 B^{ij}\rightarrow \,&\tilde{B}^{ij}=U^{i}{}_{k}B^{kl}(U^{T})_{l}{}^{j}\nonumber\\
 \epsilon^{i}\rightarrow \,&\tilde{\epsilon}^i=U^{i}{}_{j} \epsilon^{j},\, \epsilon_{i}\rightarrow\tilde{\epsilon}_i=\epsilon_{j}(U^{-1})^{j}{}_{i}
\end{align}
where the Majorana condition ($C$ being the charge conjugation matrix)
\begin{align}
    \epsilon^{i}=i\gamma^0C^{-1}\left(\epsilon_i\right)^*
\end{align}
necessitates that $U$ is a unitary matrix. We will use this freedom in the next subsection to go to a basis where the matrices $A^{ij}$ and $B^{ij}$ becomes block diagonal. In this choice of basis, it becomes easier to find the non-trivial supersymmetries preserved by the constant moduli solution as we will discuss in the next subsection.
\subsection{Supersymmetry preserved by the constant moduli solution}
In order to obtain the non-trivial supersymmetries preserved by the constant moduli solution, it is useful to block diagonalize the matrix $A$. This can be done by transforming it with a matrix $U$ \eqref{transf} made out of the orthonormal basis vectors spanning the two dimensional kernel of $A$ ($Ker(A))$ and orthonormal basis vectors spanning the two dimensional subspace orthogonal to the kernel ($Ker(A)^{\perp}$). This requirement gives us the following $U$: 
\begin{align}\label{su4matrix}
    U=\begin{pmatrix}
       \frac{P^5}{\sqrt{N_1}}&-\frac{P^3}{\sqrt{N_1}}&0&-\frac{P^4}{\sqrt{N_1}}\\\\
       \frac{-\mathcal{A}P^5}{\sqrt{N_2}}&\frac{-P^1+\mathcal{A}P^3}{\sqrt{N_2}}&\frac{P^5}{\sqrt{N_2}}&\frac{P^6+\mathcal{A}P^4}{\sqrt{N_2}}\\\\
       \frac{P^{3*}}{\sqrt{M_1}}&  \frac{P^{5*}}{\sqrt{M_1}}&  \frac{P^{1*}}{\sqrt{M_1}}&0\\\\
       \frac{P^{4*}-\mathcal{B}P^{3*}}{\sqrt{M_2}}&\frac{-\mathcal{B}P^{5*}}{\sqrt{M_2}}& \frac{-P^{6*}-\mathcal{B}P^{1*}}{\sqrt{M_2}}& \frac{P^{5*}}{\sqrt{M_2}}
    \end{pmatrix}
\end{align}
where,
 \begin{align}
    N_1=\,&|P^5|^2+|P^3|^2+|P^4|^2\\
    \mathcal{A}=\,&\frac{1}{{N_1}}\Big({P^{3*}P^1-P^{4*}P^6}\Big)\\
    N_2=\,&-|\mathcal{A}|^2 N_1+|P^1|^2+|P^6|^2+|P^5|^2 \\
    M_1=\,&|P^5|^2+ |P^3|^2+|P^1|^2\\
    \mathcal{B}=\,&\frac{1}{M_1}\Big(P^3P^{4*}-P^{1}P^{6*}\Big)\\
    M_2=\,& -|\mathcal{B}|^2 M_1+|P^4|^2+|P^6|^2 +|P^5|^2
\end{align}
The transformed matrix $\tilde{A}$ takes the following block diagonal form:
\begin{align}
      \tilde{A}&=\begin{pmatrix}
    0&0&0&0\\
   0&0&0&0\\
0&0&0&\lambda_1\\
0&0&-\lambda_1&0
\end{pmatrix}
\end{align}
where, the skew eigenvalue
\begin{align}
    \lambda_1&=\frac{P^{5*}}{|P^5|}\sqrt{|P^{1}|^2+|P^{2}|^2+|P^{3}|^2+|P^{4}|^2+|P^{5}|^2+|P^{6}|^2}
\end{align}
Using the Pfaffian condition \eqref{Pfian}, one can also check that $Ker(A)=Ker(B)^{\perp}$ and $Ker(A)^{\perp}=Ker(B)$. This implies that the same $U$ also block diagnozies the matrix $B$ as shown below:
\begin{align}
\tilde{B}&=\begin{pmatrix}
    0&\lambda_2&0&0\\
   -\lambda_2&0&0&0\\
0&0&0&0\\
0&0&0&0
\end{pmatrix}
\end{align}
where the skew-eigenvalue
\begin{align}\label{lambda2def}
    \lambda_2&=\frac{P^{5}}{|P^5|}\sqrt{|P^{1}|^2+|P^{2}|^2+|P^{3}|^2+|P^{4}|^2+|P^{5}|^2+|P^{6}|^2}
\end{align}
The transformed spinors $\tilde{\epsilon}^i$ are given as:
  \begin{align}
    \tilde{\epsilon}_{(0)}^{1}=\,&\Bigg(\frac{P^{5}}{\sqrt{N_1}}\epsilon_{(0)}^{1}-\frac{P^{3}}{\sqrt{N_1}}\epsilon_{(0)}^{2}-\frac{P^{4}}{\sqrt{N_1}}\epsilon_{(0)}^{4}\Bigg)\\
     \tilde{\epsilon}_{(0)}^{2}=\,&\Bigg(-\frac{\mathcal{A}^{}P^{5}}{\sqrt{N_2}}\epsilon_{(0)}^{1}+\frac{\mathcal{A}P^{3*}-P^{1}}{\sqrt{N_2}}\epsilon_{(0)}^{2}\nonumber\\ \,&+\frac{P^{5}}{\sqrt{N_2}}\epsilon_{(0)}^{3}+\frac{\mathcal{A}P^{4}+P^{6}}{\sqrt{N_2}}\epsilon_{(0)}^{4}\Bigg)\\
     \tilde{\epsilon}_{(0)}^{3}=\,&\Bigg(\frac{P^{3*}}{\sqrt{M_1}}\epsilon_{(0)}^{1}+\frac{P^{5*}}{\sqrt{M_1}}\epsilon_{(0)}^{2}+\frac{P^{1*}}{\sqrt{M_1}}\epsilon_{(0)}^{3}\Bigg)\\
     \tilde{\epsilon}_{(0)}^{4}=\,&\Bigg(\frac{P^{4*}-\mathcal{B}P^{3*}}{\sqrt{M_2}}\epsilon_{(0)}^{1}-\frac{\mathcal{B}^{}P^{5*}}{\sqrt{M_2}}\epsilon_{(0)}^{2}\nonumber\\ \,&-\frac{P^{6*}+\mathcal{B}P^{1*}}{\sqrt{M_2}}\epsilon_{(0)}^{3}+\frac{P^{5*}}{\sqrt{M_2}}\epsilon_{(0)}^{4}\Bigg)
\end{align}
The equation $\tilde{A}^{ij}\tilde{\epsilon}_{(0)j}=0$ implies
\begin{align}\label{34}
    \tilde{\epsilon}_{(0)}^{3}=\tilde{\epsilon}_{(0)}^{4}=\tilde{\epsilon}_{(0)3}=\tilde{\epsilon}_{(0)4}=0
\end{align}
We are left with two supersymmetry parameters $\tilde{\epsilon}_{(0)1}$ and $\tilde{\epsilon}_{(0)2}$, which we will obtain by solving the equation
\begin{align}\label{tildebeq}
   \tilde{B}^{ij}\tilde{\epsilon}_{(0)j}=\,\frac{r_{H}}{\sqrt{\chi_{0}}} \gamma_{0}\tilde{\epsilon}_{(0)}^{i} 
\end{align}. 
A non trivial solution will exist provided 

% Using the Pfaffian condition \eqref{Pfian}, one can show that $u_1,u_2\in Ker(B)$ i.e.,
%  \begin{align}
%    B.u_1=B.u_2=0
%  \end{align}
% and $v_1,v_2\in Ker(B)^{\perp}$.
 
%Once again, the phase factor gets cancelled from both sides of the equation \eqref{conkilling2}. Then   
%the condition $\tilde{B}^{ij}\tilde{\epsilon}_{(0)j}=\frac{r_{H}}{\sqrt{\chi_{0}}} \gamma_{0}\tilde{\epsilon}_{(0)}^{i}$ gives us 
%\begin{align}
   % \lambda_2\tilde{\epsilon}_{(0)2}=\,&\frac{r_{H}}{\sqrt{\chi_0}}\gamma_{0}\tilde{\epsilon}^{1}_{(0)}\\
   % \lambda_2\tilde{\epsilon}_{(0)1}=\,&-\frac{r_{H}}{\sqrt{\chi_0}}\gamma_{0}\tilde{\epsilon}^{2}_{(0)}
%\end{align}
%From these two equations, we can derive the following  
%\begin{align}
   % (\epsilon_{(0)2}^{*})^{*}=\frac{r_{H}^2}{\chi_{0} |\lambda_2|^2}\epsilon_{(0)2}
%\end{align}
% The consistency condition for non zero $\tilde{\epsilon}_{(0)1}$ and $\tilde{\epsilon}_{(0)2}$ is 
\begin{align}
    |\lambda_2|^2=\frac{r_{H}^2}{\chi_{0}}
\end{align}
Indeed, we can check that this constraint is satisfied by $\lambda_2$ given in \eqref{lambda2def}:
\begin{align}
    |\lambda_2|^2\,&=|P^{1}|^2+|P^{2}|^2+|P^{3}|^2+|P^{4}|^2+|P^{5}|^2+|P^{6}|^2\nonumber\\
    \,&=\frac{1}{4}\mathcal{Q}^{I}\bar{Q}^{J}\delta_{IJ}=-p^2=\frac{r_{H}^2}{\chi_{0}}
\end{align}
Hence, the solution of \eqref{tildebeq} gives us non-trivial  $\tilde{\epsilon}_{(0)1}$ and $\tilde{\epsilon}_{(0)2}$ which are related by a phase 
\begin{align}\label{12}
  \tilde{\epsilon}_{(0)1}=e^{i\gamma}\gamma_{0}\tilde{\epsilon}_{(0)}^{2}
\end{align}

\textit{Thus we have shown that an extremal black hole solution with constant moduli always preserves $1/4$-supersymmetry for generic charge configuration satisfying $p^2q^2>(q.p)^2$ in pure $\mathcal{N}=4$ supergravity.}
\section{Discussions}\label{conclusion}
In this work, we have studied the attractor behaviour of extremal black holes in pure $\mathcal{N}=4$ leading-order Poincaré supergravity using the black hole potential approach. We performed a perturbative expansion around the constant moduli solution and showed that attractor behaviour works to all orders in perturbation theory. Furthermore, we used the results of first-order perturbation theory to set the boundary condition near the horizon and numerically solved the exact field equations, explicitly showing the attractor behaviour. 
We also analysed the supersymmetry of the constant moduli solution. In order to do that, we solved the Killing spinor equations by exploiting the superconformal framework and showed that the constant moduli solutions are always $1/4$-BPS.

A classification of BPS and non-BPS black holes for various supergravity theories have been discussed in \cite{Andrianopoli:1997pn, Andrianopoli:2006ub,Bellucci:2006xz, Cerchiai:2009pi, Ceresole:2010hq}. The classification is based on the orbits of the corresponding electro-magnetic duality groups of the theory which are characterised by the duality invariant combination of the electric and magnetic charges. Depending on the value of quatric invariant $I_{4}=p^2q^2-(p.q)^2$, there exists three branches of solutions in  $\mathcal{N}=4$ supergravity: the $1/4$-BPS branch ($I_{4}>0$) which exists in both pure and matter coupled theories, two inequivalent non-BPS branch with $I_{4}>0$ and $I_{4}<0$ (only exists in matter coupled theory). Based on this, we expect that the ``non-constant moduli'' solutions belonging to the family of attractor solutions given in {Figures \ref{fig:axion1}, \ref{fig:dilaton1}, \ref{fig:ghj}}, which are characterised by $I_{4}>0$ are also $1/4$-BPS solutions, although we have  not explicitly analysed the supersymmetry of these non-constant moduli solutions. We also expect this to be true since all the ``non-constant moduli'' solutions are continuously connected to the constant moduli solutions which we have explicitly shown to be $1/4$-BPS. In this sense, we can infer that all the attractor solutions in pure $\mathcal{N}=4$ supergravity is expected to be $1/4$-BPS. 

The constraints \eqref{34} and \eqref{12} suggest a possible structure of projection conditions on the Killing spinors, which may be used to construct more general $1/4$-BPS solutions of $\mathcal{N}=4$ supergravity. Interestingly, constraints of a similar type have appeared in the study of general stationary axion-dilaton solutions in four-dimensional two-derivative pure $\mathcal{N}=4$ supergravity, namely the so-called SWIP solutions \cite{Bergshoeff:1996gg, Tod:1995jf, Bellorn2005AllTS}. These solutions are generically $1/4$-BPS and can be viewed as the $\mathcal{N}=4$ generalisation of the standard Israel--Wilson--Perj\'es (IWP) solutions in $\mathcal{N}=2$ supergravity. The supersymmetry of the SWIP solutions has been analysed in a setting where four of the six vector fields of the pure $\mathcal{N}=4$ theory are truncated out. We would like to understand if an analysis of the most general $1/4$-BPS solutions in pure $\NN=4$ supergravity using the projection conditions of the type discussed in \eqref{34} and \eqref{12} would be able to recover the SWIP solutions and even study further generalizations of it to matter-coupled theories.

The study of supersymmetric solutions in higher derivative $\NN=4$ supergravity is useful and our motivation for such a study stems from what we already know in $\NN=2$ supergravity. The fully supersymmetric solutions of $\mathcal{N}=2$ supergravity have played a crucial role in finding higher derivative corrections to black hole entropy. In \cite{LopesCardoso:2000qm}, a set of constraints was obtained from the Killing spinor equations, which are necessary conditions for half-supersymmetric solutions. Although, the explicit form of the most general solutions satisfying these constraints remains unknown, nevertheless these general constraints have played a crucial role in the recent computation of gravitational index of supersymmetric extremal black holes in higher derivative $\mathcal{N}=2$ supergravity \cite{Hegde:2024bmb}. In higher-derivative $\mathcal{N}=4$ supergravity, fully supersymmetric solutions have recently been analysed using the superconformal framework \cite{Bhattacharjee:2025qro}. This analysis was done for a class of higher-derivative actions in $\NN=4$ supergravity that contains terms involving supersymmetric completion of Weyl-squared term, where the Lagrangian takes the following form
\begin{align}
    e^{-1}\mathcal{L}=-\frac{1}{2}R-\mathcal{H}(\tau)\Big(\frac{1}{2}C_{\mu\nu\rho\lambda}C^{\mu\nu\rho\lambda}+\cdots\Big),
\end{align}
where $\mathcal{H}(\tau)$ is a holomorphic function of the axion-dilaton moduli. In this class of theories, flat Minkowski spacetime is the only fully supersymmetric solution.\footnote{A general criterion for ungauged two derivative supergravity to admit non-flat fully supersymmetric was studied in \cite{Louis:2016tnz}.} This stands in contrast to higher-derivative $\mathcal{N}=2$ supergravity, where geometries such as $AdS_2 \times S^2$ are also fully supersymmetric and arise as near-horizon limits of extremal black holes which typically preserve $1/2$ of the supersymmetries. 
In the $\mathcal{N}=4$ case, $AdS_2 \times S^2$ geometries are not fully supersymmetric as seen in the analysis of \cite{Bhattacharjee:2025qro}. Hence, they are expected to be $1/2$-BPS. Such a geometry would also arise in the near-horizon limit of $1/4$-BPS extremal black holes. It would therefore be interesting to find the constraints satisfied by general $1/2$-BPS and $1/4$-BPS solutions in higher-derivative $\NN=4$ supergravity theories. Such an analysis may allow for a  derivation of the entropy formula of $1/4$-BPS black holes directly within $\mathcal{N}=4$ supergravity, expressed in terms of the holomorphic function $\mathcal{H}(\tau)$. Furthermore, this can be useful for studying recent developments of computing supersymmetric index in the context of $\mathcal{N}=4$ supergravity. We hope to return to these questions in future work.

\acknowledgments
We thank Dileep Jatkar and Anirban Basu for discussions. AB thanks Harish-Chandra Research Institute (HRI), Prayagraj, where the initial part of this project was presented. AB and BS thank Harish-Chandra Research Institute (HRI), Prayagraj, for its hospitality during the initial stages of this project. AB also thanks the Asian Winter School (AWS) 2026, IISER Bhopal, for its hospitality during the final stages of this project. This work is partially supported by ANRF MATRICS Advanced Research Grant ANRF/ARGM/2025/000640/TS, Government of India.

\appendix
\section{Notations and Conventions}\label{conv}
The gamma matrices satisfy
\begin{align}
    \{\gamma^a, \gamma^b\}=\,&2\eta^{ab}\mathbb{I},\\
        \gamma ^a\gamma^b=\,&\eta^{ab}\mathbb{I}+\gamma^{ab},~~\gamma^{ab}=\frac{1}{2}(\gamma^a\gamma^b-\gamma^b\gamma^a),
\end{align}
The four-dimensional Levi-Civita is defined as 
\begin{align}
    \varepsilon^{0123}=\,&i,~~\varepsilon_{0123}=-i.\\
    \delta^{abcd}_{efgh}=\,&4! \delta^a_{[e}\dots \delta^c_{h]}, ~~\varepsilon^{abcd}\varepsilon_{efgh}=\delta^{abcd}_{efgh},
\end{align}
Useful gamma matrix identities
\begin{align}
    \gamma_5=\,&\frac{1}{4!}\varepsilon_{abcd}\gamma^a\gamma^b\gamma^c\gamma^d, \\
    \{\gamma_{ab}, \gamma_c\}=\,&2\varepsilon_{abcd}\gamma_5 \gamma^d, \\ [\gamma_{ab}, \gamma_c]=\,&4 \gamma_{[a}\eta_{b]c}.\\
        [\gamma_{ab}, \gamma^{cd}]=\,&8\gamma_{[a}{}^{[d}\delta_{b]}{}^{c]},\\
    \{\gamma_{ab}, \gamma^{cd}\}=\,&-4\delta_{[a}^{c}\delta_{b]}^{d}+2\varepsilon_{ab}{}^{cd}\gamma_{5}.
\end{align}
The dual of an antisymmetric tensor $F_{ab}$ 
\begin{align}
\tilde{F}_{ab}=\frac{1}{2}{\varepsilon}_{abcd}F^{cd}
\end{align}
The self-dual and antiself-dual part: 
\begin{align}
    F_{ab}^{\pm}=\frac{1}{2}(F_{ab}\pm \tilde{F}_{ab})
\end{align}
\section{Convergence of the perturbation expansion}\label{convergence}
To establish convergence of the series, we bound the coefficients.
Assume that there exist positive constants $A$ and $B$ such that
\begin{align}
|h_k| \le A B^k,\, \,
|c_k| \le A B^k, \,\,
|d_k| \le A B^k ,\,\,
|f_k| \le A B^k
\end{align}
We consider the series \eqref{bseries}. The coefficients are 
\begin{align}
h_n =
-\frac{T_n - \tilde T_n}{4\chi_0^2\,n(n-1)},
\end{align}
where
\begin{align}
T_n &= \sum_{\substack{l_1+l_2+s_1=n \\ s_1>1}}
4\,h_{s_1}\,s_1(s_1-1)\,d_{l_1}d_{l_2},\\
\tilde T_n &= \sum_{\substack{l_1+l_2+s_1=n \\ l_1,l_2\ge1}}
h_{s_1}\,(c_{l_1}c_{l_2}+d_{l_1}d_{l_2}).
\end{align}
Each term in $T_n$ satisfies
\begin{align}
|h_{s_1} d_{l_1} d_{l_2}|
\le A^3 B^{s_1+l_1+l_2}.
\end{align}
Using the constraint $l_1+l_2+s_1=n$, this becomes
\begin{align}
|h_{s_1} d_{l_1} d_{l_2}| \le A^3 B^n .
\end{align}

Since $s_1< n$, we also have
\begin{align}
s_1(s_1-1) < n^2 .
\end{align}

Next we estimate the number of terms in the sum. The constraint
\begin{align}
l_1 + l_2 + s_1 = n, \qquad s_1 > 1
\end{align}
can be rewritten by defining
\begin{align}
s_1' = s_1-2 ,
\end{align}
which gives
\begin{align}
l_1 + l_2 + s_1' = n-2, \qquad l_1,l_2,s_1' \ge 0 .
\end{align}
The number of non-negative integer solutions of
\begin{align}
x_1+x_2+x_3=N
\end{align}
is given by the stars-and-bars formula
\begin{align}
\binom{N+2}{2}.
\end{align}
Thus, the number of solutions is
\begin{align}
\binom{n}{2}
=
\frac{(n)(n-1)}{2}
\sim \mathcal{O}(n^2).
\end{align}

Combining these estimates, we obtain
\begin{align}
|T_n|
&\le
\sum_{\substack{l_1+l_2+s_1=n}} 
4\,|h_{s_1}|\,s_1(s_1-1)\,|d_{l_1}|\,|d_{l_2}| \\
&\le
C_1 A^3 B^n n^4
\end{align}
for some constant $C_1>0$.

Similarly, we can bound $\tilde{T}_{n}$
\begin{align}
|\tilde T_n|
\le
C_2 A^3 B^n n^2
\end{align}
for some constant $C_2>0$.

Using the recursion relation, we obtain
\begin{align}
|h_n|
\le
\frac{|T_n|+|\tilde T_n|}
{4\chi_0^2\,n(n-1)}.
\end{align}

This in turn bounds $h_{n}$ as 
\begin{align}
|h_n|
\le
C\,A^3 B^n n^2
\end{align}
for some constant $C>0$. Now the radius of convergence is given by the Cauchy-Hadamard fornmula
\begin{align}
R^{-1}\equiv\limsup_{n\to\infty} |h_n|^{1/n} \le B .
\end{align}

Therefore the series
\begin{align}
b(r) = \sum_{n=0}^{\infty} h_n x^n ,
\qquad
x = 1-\frac{r_H}{r},
\end{align}
has a non-zero radius of convergence
\begin{align}
R \ge \frac{1}{B}.
\end{align}
 One can show convergence for other series also by applying similar method. 
\bibliography{bibliography}
\bibliographystyle{JHEP}% Produces the bibliography via BibTeX.
\end{document}